\documentclass[sn-mathphys-ay]{sn-jnl}

\usepackage{graphicx}%
\usepackage{multirow}%
\usepackage{amsmath,amssymb,amsfonts}%
\usepackage{amsthm}%
\usepackage{mathrsfs}%
\usepackage[title]{appendix}%
\usepackage{xcolor}%
\usepackage{textcomp}%
\usepackage{manyfoot}%
\usepackage{booktabs}%
\usepackage{algorithm}%
\usepackage{algorithmicx}%
\usepackage{algpseudocode}%
\usepackage{listings}%
\usepackage{natbib}%
\usepackage{lipsum} %
\usepackage{subcaption}%
\usepackage{float}%

\begin{document}

\title[An attempt to build a dynamical catalog of Solar System co-orbitals]{An attempt to build a dynamical catalog of Solar System co-orbitals}

\author*[1]{\fnm{Nicolas} \sur{Pan}}\email{nicolas.pan@fcien.edu.uy}

\author[1]{\fnm{Tabaré} \sur{Gallardo}}\email{tabare.gallardo@fcien.edu.uy}

\affil*[1]{\orgdiv{Departamento de Astronomía}, \orgname{Facultad de Ciencias}, \orgaddress{\street{Iguá 4225}, \city{Montevideo}, \postcode{11400}, \state{Montevideo}, \country{Uruguay}}}


\abstract{

The main objective of this paper is to fully study 1:1 mean-motion resonance in the Solar System. We calculated stability points applying a resonant semi-analytic theory valid for any value of eccentricity or inclination.
The location of each equilibrium point changes as the orbital elements of an object change, which led us to map the location of them. For the case of low inclination and low eccentricity we recovered the known L4 and L5 points. The three global types of orbits for this resonance; Tadpoles, Quasi-satellites and Horseshoes, vary as a function of the orbital elements, even disappearing for some cases.
In order to build a catalog of real co-orbital objects, we filtered the NASA Horizons asteroids catalog inside the maximum resonant width and analyzed which objects are indeed in resonance. 
In total, we found 169 objects to be in co-orbital resonant motion with Solar System planets excluding Jupiter. We were able to recover all the already known objects and to confirm the resonant state of some new ones. Mercury remains to be the only planet with zero known co-orbitals and no L5 Earth Trojan has been discovered so far.
Among the interesting identified orbits we highlight the one of 2021 FV$_{1}$, a new Mars Quasi-satellite.
Despite having circulating critical angles, we found some objects to be dynamically driven by the resonance such as 2012 QR$_{50}$.}

\keywords{Celestial Mechanics, Resonances, Co-orbitals, Asteroids, Solar System}

\maketitle

\section{Introduction}
\label{sec:introduction}

\bigskip

Solar System planets have a growing population of small bodies called co-orbitals which show a great diversity of trajectories.
The co-orbital motion is possible due to the existence of stable equilibrium points around which the orbits of small bodies can oscillate.
It is known that in the circular restricted three body problem scheme there are 2 stable equilibrium points known as the Lagrangian points L4 and L5.
In terms of orbital elements, particles with small amplitude oscillations around L4 or L5 have quasi-circular and quasi-planar orbits with the planet, but the concept of equilibrium points can be generalized to arbitrary orbits within the more general framework of the 1:1 particle-planet mean-motion resonance, the most studied and well documented mean-motion resonance in Celestial Mechanics.

The variable that defines whether there is stable equilibrium in an orbital configuration in 1:1 resonance is the so-called critical angle $\sigma = \lambda - \lambda_p$, being $\lambda$ and $\lambda_p$ the mean longitudes of the small body and the planet respectively. 
The critical angle $\sigma$ must oscillate around a fixed value with a given period.
In the case of quasi-circular and quasi-coplanar orbits $\sigma$ is librating around 60$^{\circ}$ or 300$^{\circ}$ coinciding with the location of L4 and L5 respectively, that is, 60$^{\circ}$ ahead or behind the planet.
In the general case of 1:1 mean-motion resonance, the equilibrium points are not necessarily at those locations as was firstly noted by \citet{1964SAOSR.149.....S} and are not necessarily two; this research work partly aims at clarifying this frequently overlooked issue.
The most popular case in the Solar System are Jupiter Trojans, more than ten thousand objects have been found with this type of low eccentricity and low inclination orbits librating around $\sigma \sim 300^{\circ}$ or $\sigma \sim 60^{\circ}$. 
These orbits are called Trojan-like or Tadpoles (TP) due to their shape in the heliocentric frame of reference rotating with the host planet.
Apart from Tadpoles there are two other co-orbital orbits that receive the name Quasi-satellites (QS) and Horseshoes (HS). In the first one, the critical angle librates around a stable equilibrium point located at $\sigma \sim 0^{\circ}$ and in the second one $\sigma$ librates wrapping both stable equilibrium points L4 and L5 and the unstable equilibrium point L3.

Stability of co-orbitals has been deeply discussed in the literature considering various effects and scenarios. As examples, \citet{2002MNRAS.334..241B} showed the importance of secular resonances in the instability of terrestrial planet co-orbitals.
\citet{2021MNRAS.507.1640C} investigated the stability of Earth co-orbitals finding that Earth Horseshoes should be more stable than Trojans. This result was also found earlier by \citet{2019A&A...622A..97Z} and \citet{2012MNRAS.426.3051C}.
Due to the perturbations of other planets, objects sometimes change their libration center but not fully escape the resonance for certain time.
Furthermore, compound or hybrid orbits have been studied as a combination of the base orbits Tadpole, Quasi-satellite and Horseshoe \citep{1999Icar..137..293N, 2000Icar..144....1C, 2016Ap&SS.361...16D}.
\citet{2004Icar..171..102B} studied in detail HS-QS transitions for Venus and Earth co-orbitals. 
Other transition have been observed for Jupiter co-orbitals from Quasi-satellite to Tadpole orbits \citep{2012AcA....62..113W}.

Co-orbital objects have been discovered and confirmed not only with Jupiter but with other Solar System planets, with the exception of Mercury. As described by \citet{2006Icar..185...29M} it is not simple to detect such objects in inner Earth orbits. 
Thus, the absence of Mercury co-orbitals could just be an observational bias. A clear example of this is the high eccentricity distribution of Venus confirmed co-orbitals. Most of these are discovered when they reach aphelion, making observation easier.
One of the main conclusions of \citet{2006Icar..185...29M} was that the population of NEAs in coorbital motion with Venus was complete up to a magnitude of H=22. Thus, smaller objects could be found in resonance with those planets and in this work we present some new candidates.

The study of co-orbital populations may yield information about the formation of our Solar System.
Long term stability of Trojans has been widely studied in the literature. It has been proposed that some Jupiter and Neptune Trojans may be in stable orbits for at least the age of the Solar System \citep{1997Natur.385...42L, 2006MNRAS.372.1463R, 2007MNRAS.382.1324D}. These objects are called primordial Trojans.

Different conclusions have been discussed in the literature regarding the stability of co-orbital orbits. As an example, \citet{2005AJ....130.2912S} claimed that Venus Trojans could not be primordial due to the effect of secular resonances and Yarkovsky effect.
Nevertheless, a circular population of Venus Trojans has been proposed to explain Venus' Zodiacal Dust Ring by \citet{2019ApJ...873L..16P}.
It was latter shown by \citet{2021PSJ.....2..193P} that the current Venus co-orbital population likely did not originate from these proposed population.
Thus, a scenario where inner planets co-orbitals scattered from the asteroid belt remains as the main idea for these object's origin.
\citet{2014MNRAS.445.3999G} also proposed Hungarian region as a source.

Earth co-orbital population is observationally biased to Quasi-satellite objects due to the closer distances they reach.
Some studies suggested that this population's stability is highly affected by Yarkovsky drift but only focusing on near-circular and near-planar orbits \citep{2019A&A...622A..97Z, 2021MNRAS.507.1640C}.
Nonetheless, Horseshoe co-orbitals have also been discovered \citep{1998AJ....115.2604W, 2020MNRAS.496.4420K} as well as Earth Trojans \citep{2012A&A...541A.127D, 2021ApJ...922L..25H}. None of them has been shown to be stable in the long term.

With respect of Mars Trojans, \citet{2005Icar..175..397S} showed that all four objects known at the moment resided in the most stable region. Three new stable Mars Trojans were latter confirmed by \citet{2013MNRAS.432L..31D}. On the other hand, \citet{2005P&SS...53..617C} showed the existence of some transient ones.
This transient behavior appears to be a common characteristic among all these bodies. Some studies also suggest that Mars Trojans are impact ejecta from Mars rather than captured bodies from the main belt \citep{2017NatAs...1E.179P}.
Similar formation process has also been proposed for Earth's co-orbital Kamo\text{\textquoteleft}oalewa \citep{jiao2024asteroid}.

1999 RG$_{33}$ has been proposed as a Saturn co-orbital by \citet{2006Icar..184...29G}. Additionally, 2001 BL$_{41}$ has been reported to be a short-lived Quasi-satellite of such planet \citep{2001MPEC....B...44G, 2016MNRAS.462.3344D}.
\citet{2014MNRAS.437.1420H} studied the problem from a more theoretical perspective and concluded that secular resonances and the near conmensurability between the libration frequency and the great inequality are the cause of instability for Saturn co-orbitals.
Jupiter perturbations are mostly responsible for destabilizing Saturn co-orbitals.

Regarding Uranus co-orbitals, \citet{2006Icar..184...29G} reported  Crantor (2002 GO$_{9}$) as the first object co-orbital with Uranus moving in a Horseshoe orbit while the first Uranus transient Trojan was reported by \citet{2013Sci...341..994A}.
Crantor was latter studied in further detail by \citet{2013A&A...551A.114D} where they showed the instability of Crantor orbit. These authors have also reported several other Uranus co-orbitals \citep{2014MNRAS.441.2280D, 2015MNRAS.453.1288D, 2017MNRAS.467.1561D}.
Recently, the stability of these and other hypothetical Uranus companions has been studied, revealing a small probability of their stability over the age of the Solar System \citep{2020A&A...633A.153Z, 2023MNRAS.519..812W}.
Temporary Uranus Quasi-satellite has been reported by \citet{2012A&A...545L...9D} and \citet{2012A&A...547L...2D}.

Neptune co-orbital population is the second largest known in our Solar System. Most of the discovered objects are located in L4, this asymmetry is shared with Jupiter Trojans. Stability has been a matter of many studies.
\citet{2014Icar..243..287D} showed that the escape rates are similar for L4 and L5 fictitious Jupiter Trojans in the age of the Solar System while real objects show a difference in future chaotic diffusion.
Several articles have studied the influence of migration processes in the early Solar System, showing their effects on the current configuration of Trojan populations \citep{2007MNRAS.382.1324D, 2009MNRAS.398.1715L, 2010ApJ...723L.233S, 2009AJ....137.5003N}.
Some of these objects have been reported as potentially primordial ones \citep{2004MNRAS.347..833B} while other articles have claimed Neptune Trojans as a source of centaurs \citep{2010MNRAS.405...49H}. 

To sum up, the current catalogue of Solar System planets co-orbitals is highly biased and dynamical studies can give us clues about the real population. Most studies and surveys have been historically biased to low eccentricity and low inclination orbits due to the restrictions in theory developments and ease of observations. In this article, we show that using a resonant semi-analytic theory combined with numerical integrations can help us further identify the dynamics of co-orbital candidates without any restrictions in $(e,i)$.

The work is structured as follows:
In Section 2 we present the semi-analytical theory we used and describe the numerical integrations performed. In Section 3 we show the results of the model as well as the objects found to be in co-orbital motion in the Solar System. In Section 4 we describe the conclusions of this work.

\section{Methods}
\label{sec:methods}

As discussed before, co-orbitals are in 1:1 mean-motion resonance with their host planet. Several approaches have been developed to study orbital evolution under the influence of mean motion resonances.
Both analytical and numerical methods have their own advantages and drawbacks. Most analytical methods rely on assumptions, including constraints on particle eccentricity, inclination, and the orbital eccentricity of the host planet.
Several formalisms have been developed for this resonance, limited to certain values of eccentricities and inclinations. \citep{2023MNRAS.522.2821T, 2023AcASn..64...50L}
A complete theory for co-orbital motion valid for any values of ($e,i$) including secular evolution when adiabatic invariance is valid, can be found in \citet{2002CeMDA..82..323N}.
On the other hand, purely numerical studies do not allow us to have a deeper understanding of the underlying physics behind the dynamics. Semi-analytical models offer us a middle point. In this paper we will use the resonance model developed by \citet{2020CeMDA.132....9G} This model is simpler than analytical approaches and only provides instantaneous properties assuming that the orbital elements ($e, i, \omega, \Omega$) are fixed. We can follow the changes in the resonance topology as the orbital elements evolve secularly, but we can not predict how the evolution will be in the long-term.
The Hamiltonian of a particle in 1:1 mean-motion resonance is

\begin{equation}
    \mathcal{H} (a, \sigma) = - \frac{\mu}{2a} - n_p \sqrt{\mu a} - \mathcal{R} (a_0, \sigma)
\end{equation}
where $\mu = GM_{\star}$ being $G$ the gravitational constant, $M_{\star}$ the mass of the star, $a$ is the semi-major axis of the particle, 
$n_p$ is the mean motion of the planet and $a_0$ is the nominal value for the resonance.
The first term is due to keplerian motion, the second arises from the passage to the extended phase space, and the third term, $\mathcal{R}$, represents the resonant disturbing function calculated numerically which depends on the fixed orbital parameters ($e, i, \omega, \Omega$) of the particle.
 Additionally, the model provides several resonance parameters, including the period of libration for small amplitude librations, the center of libration, and the width of the resonance.
These depend on the orbital configuration as well as the mass of the host star and the orbit of the perturbing planet.
For the particular 1:1 mean-motion resonance we can link this resonant perspective with Lagrange equilibrium points.
In this model, certain resonance libration centers correspond to the Lagrangian points L4 and L5, which are here identified as minima of the resonant function.

Since we are interested in studying the 1:1 mean-motion resonance with Solar System planets, we must calculate the resonance parameters for each of them. This calculation must consider the dependency of these parameters on the orbital elements of the asteroid.
Given the set of variables that define the star and planet $(M_{\star},a_{p},e_{p},m_{p})$, the semi-analytic model calculates a resonant width that depends on the asteroid's orbital elements $(e,i,\omega,\Omega)$. Here, the inclination $i$ is referenced to the reference plane in which the planet is assumed to be in.
As the orbital elements of an asteroid evolve, the resonant width also changes. To account for all possible evolutionary paths an asteroid could have taken, we determined the maximum resonance width as described below.
As a reference, Table \ref{tab:resonance_params} displays some resonance properties used in this work, including the maximum width.
This width was calculated by sampling the space of orbital elements $e \in (0,1), i \in (0^{\circ},180^{\circ}), \omega \in (0^{\circ},360^{\circ})$ and $\Omega \in (0^{\circ},360^{\circ})$ and finding the maximum value among all grid points within the specified parameter space.
For the calculation of the maximum width, $R_{Hill}$ factor was set to 3, corresponding to the maximum width of stable librations, as described by \citet{2020CeMDA.132....9G}.
The libration periods are given just as reference and they were calculated for all the planets considering their real eccentricities and a fictitious asteroid with orbital elements ($e = 0.2$, $i = 15 ^{\circ}$, $\omega = 60^{\circ}$, $\Omega = 0^{\circ}$).

\begin{table}[h!]
	\centering
	\begin{tabular}{cccccc}
	    Planet & $a_{0}$ [au] & Max. Width [au] & QS Period [yr] & T Period [yr] & N$^\circ$ candidates\\
          \hline
            Venus   & 0.723 & 0.016 & 75 & 150 & 88 \\
            Earth   & 1.000 & 0.023 & 105 & 220 & 657 \\
	    Mars    & 1.524 & 0.015 & 517 & 1258 & 687 \\
		Jupiter & 5.202 & 0.854 & 64 & 147 & 14484 \\
		Saturn  & 9.551 & 1.099 & 286 & 669 & 114 \\
		Uranus  & 19.176 & 1.235 & 2115 & 2870 & 87 \\
            Neptune & 30.099 & 2.000 & 4382 & 8772 & 107 \\
        \end{tabular}
        \caption{Nominal semi-major axis for resonance 1:1 with each planet, maximum resonance width for each planet, libration period for a Quasi-satellite and a Tadpole orbit for an asteroid with orbital elements ($e = 0.2$, $i = 15 ^{\circ}$, $\omega = 60^{\circ}$, $\Omega = 0^{\circ}$). Last column shows the number of objects taken from the catalog to be inside two resonance widths. Data obtained from the NASA Horizons system in July 2024.}
        \label{tab:resonance_params}
\end{table}

As previously stated, our objective was to build a catalog of Solar System co-orbitals including new candidates and to study their dynamics.
In order to accomplish this, we first searched in NASA Horizons database for objects with semi-major axis inside two maximum resonance widths for each planet, without imposing restrictions on any other orbital parameter. Orbital data was obtained from the Horizons system in July 2024.
We consider that this criterion would allow us to capture all resonant objects minimizing false positives. Any false positives will be discarded later. The resonance maximum width and the number of co-orbital resonant candidates are shown in Table \ref{tab:resonance_params}.
Given the enormous size of Jupiter's co-orbital population, we excluded those objects from this study, focusing instead on the lesser-known populations of other planets.
We acknowledged that not all orbits are well-determined, hence some object orbits may be fictitious. However, these orbits provide insight into potential unseen co-orbital populations resulting from biased observation campaigns. After we carried out the query we cross-checked the results with already known co-orbitals reported in the literature, as detailed in section \ref{sec:introduction}. 

The simplest method to determine if an object is in resonance or not is to calculate the critical angle $\sigma$ and check whether it librates around one of the equilibrium values with the libration period predicted by the theory. 
We will show later that this criterion leaves out some objects that are not strictly on resonance but their dynamics is strongly determined by it.
In addition, we know that the semi-major axis must evolve close to the mean semi-major axis of the planet and must oscillate with the same period as $\sigma$.
Likewise, the movement of a particle must be restricted to a surface of constant $\mathcal{H}$. Joining the two criteria, the best choice is to follow the trajectory in the phase space ($\sigma$, $a$) and to compare with the level curves of the Hamiltonian given by the semi-analytic theory.

To verify the resonant behavior of each co-orbital candidate, we integrated each one over several libration periods and plotted the time evolution of $\sigma$ and the particle's trayectory in ($\sigma$, $a$) plane, comparing the results of the numerical integrations with the Hamiltonian level curves.
This process enabled us to discard non-resonant objects and classify the resonant ones according to their orbit type.
We eliminated all candidates that did not show oscillation for at least one libration period. Objects with shorter librations may appear as ``jumping'' in the final list if they show recurring librations but changing the libration center among the equilibrium points.
As we discuss latter, the actual trajectory of an asteroid does not always strictly follows the level curves due to perturbations from other planets not involved in the resonant configuration.
Despite considering the maximum width of the resonance, it is important to emphasize that this work focuses on constructing an instantaneous catalog. 
The evolution of an orbit is influenced not only by resonant perturbations but also by other factors that can destabilize the current orbit over secular timescales. Secondary resonances, secular resonances, and non-gravitational forces were not taken into account in this analysis as the timescales in which they affect are longer than the instantaneous time-span we focused on.
All numerical integrations were performed starting at epoch JD$2460400.5$ using ias15 integrator from REBOUND package \citep{rebound, reboundias15} as it effectively handles close encounters between planets and the massless particles integrated in this study. Importantly, all integrations maintained a relative energy error below 10$^{-14}$.
EVORB integrator was used to check the final results \citep{2002Icar..159..358F}.
In all numerical integrations, perturbations from all Solar System planets were taken into account.

\section{Results}

\subsection{Equilibrium points and topology of $\mathcal{H}$ for arbitrary $(e,i)$}
\begin{figure}[h!]
  \centering
  \begin{tabular}{cc}
    \includegraphics[width=0.45\linewidth]{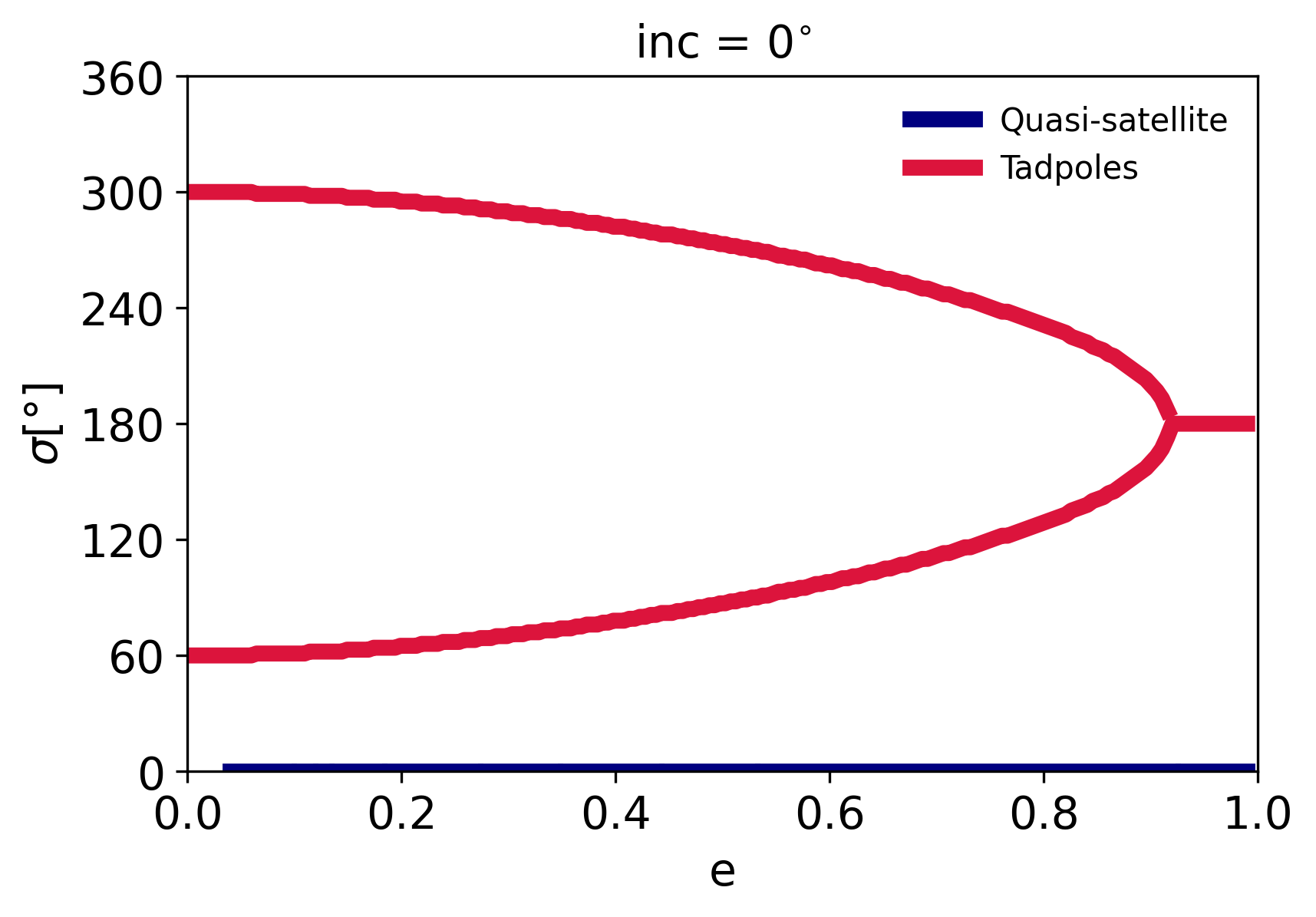} & \includegraphics[width=0.45\linewidth]{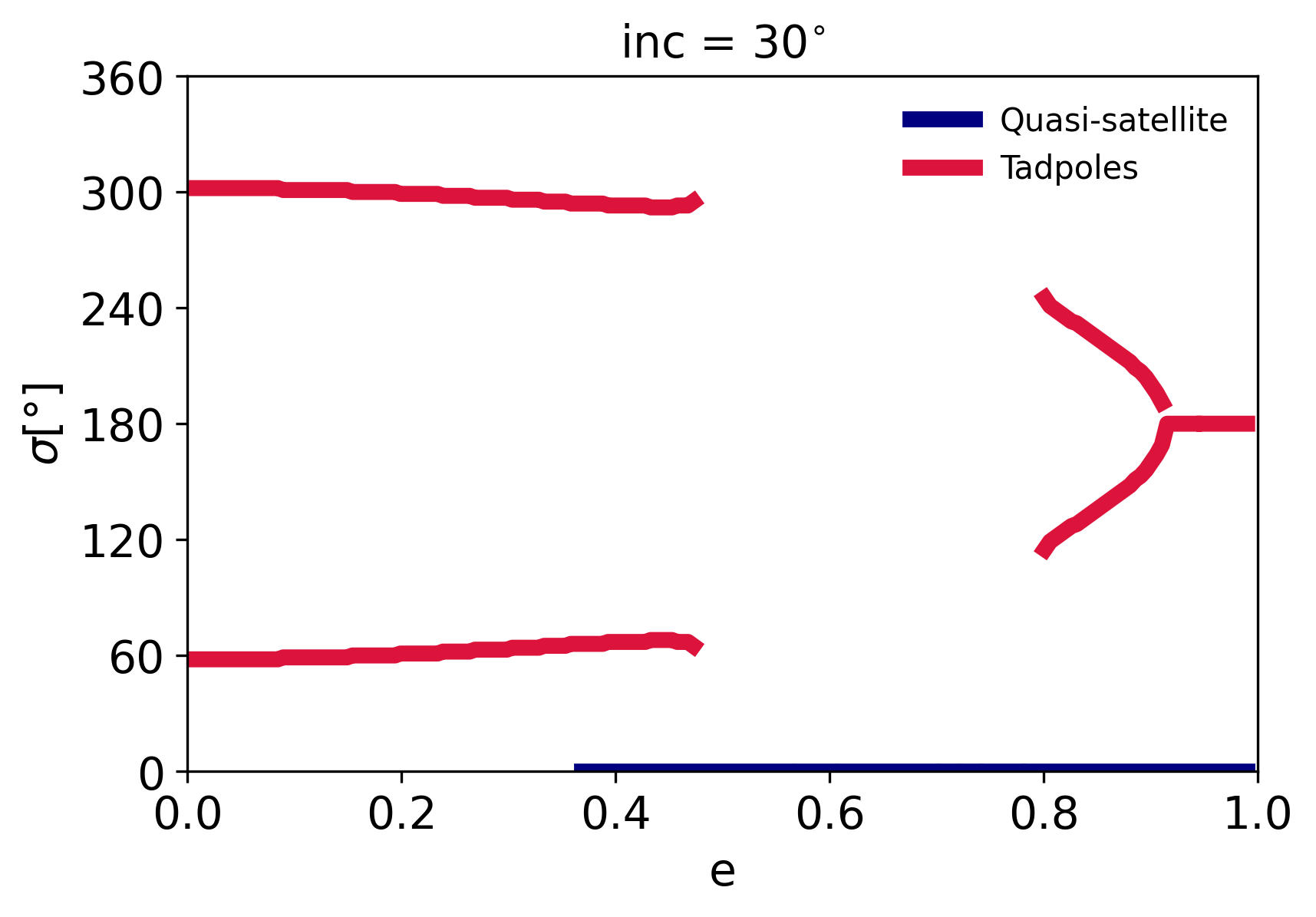} \\
    \includegraphics[width=0.45\linewidth]{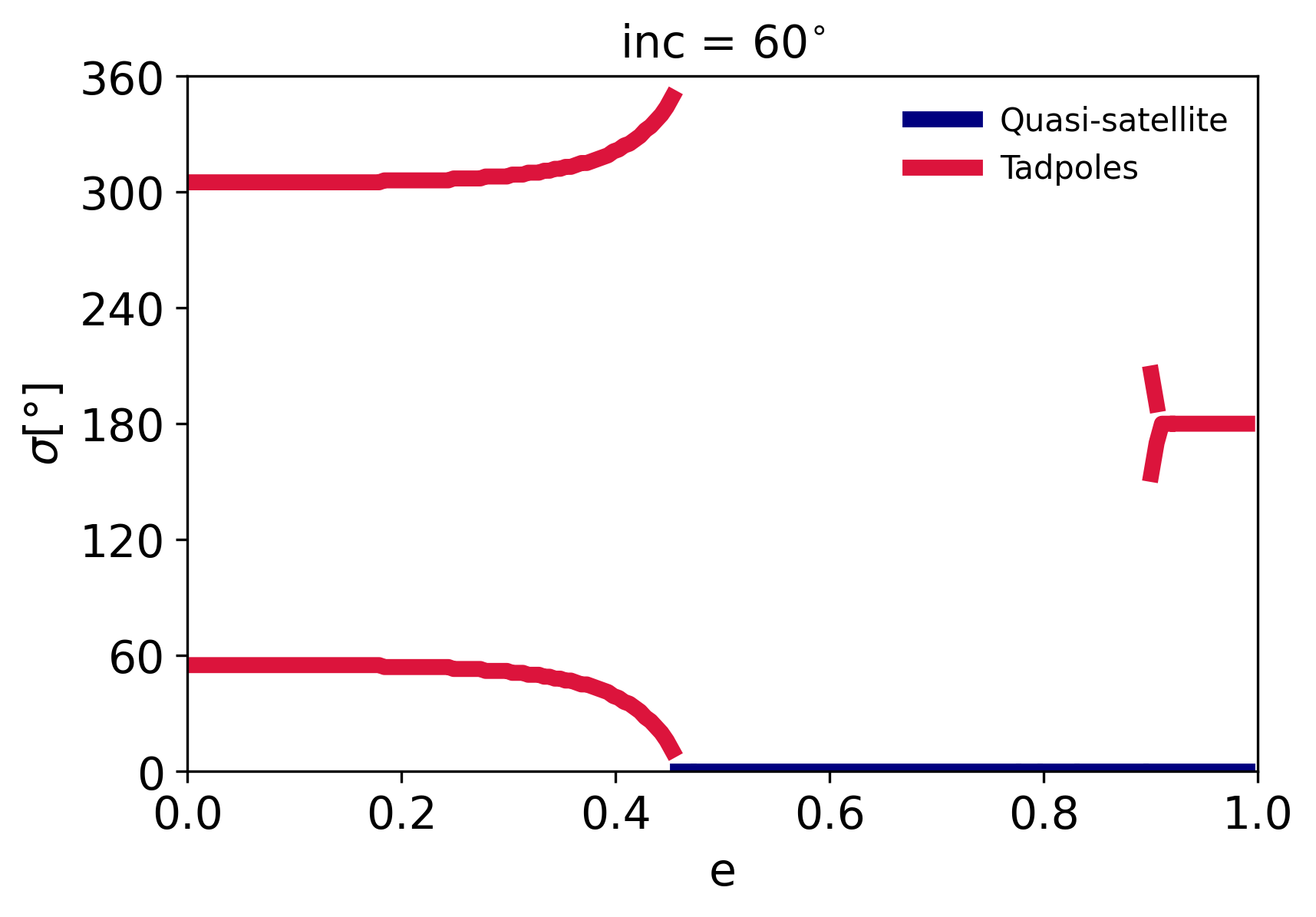} & \includegraphics[width=0.45\linewidth]{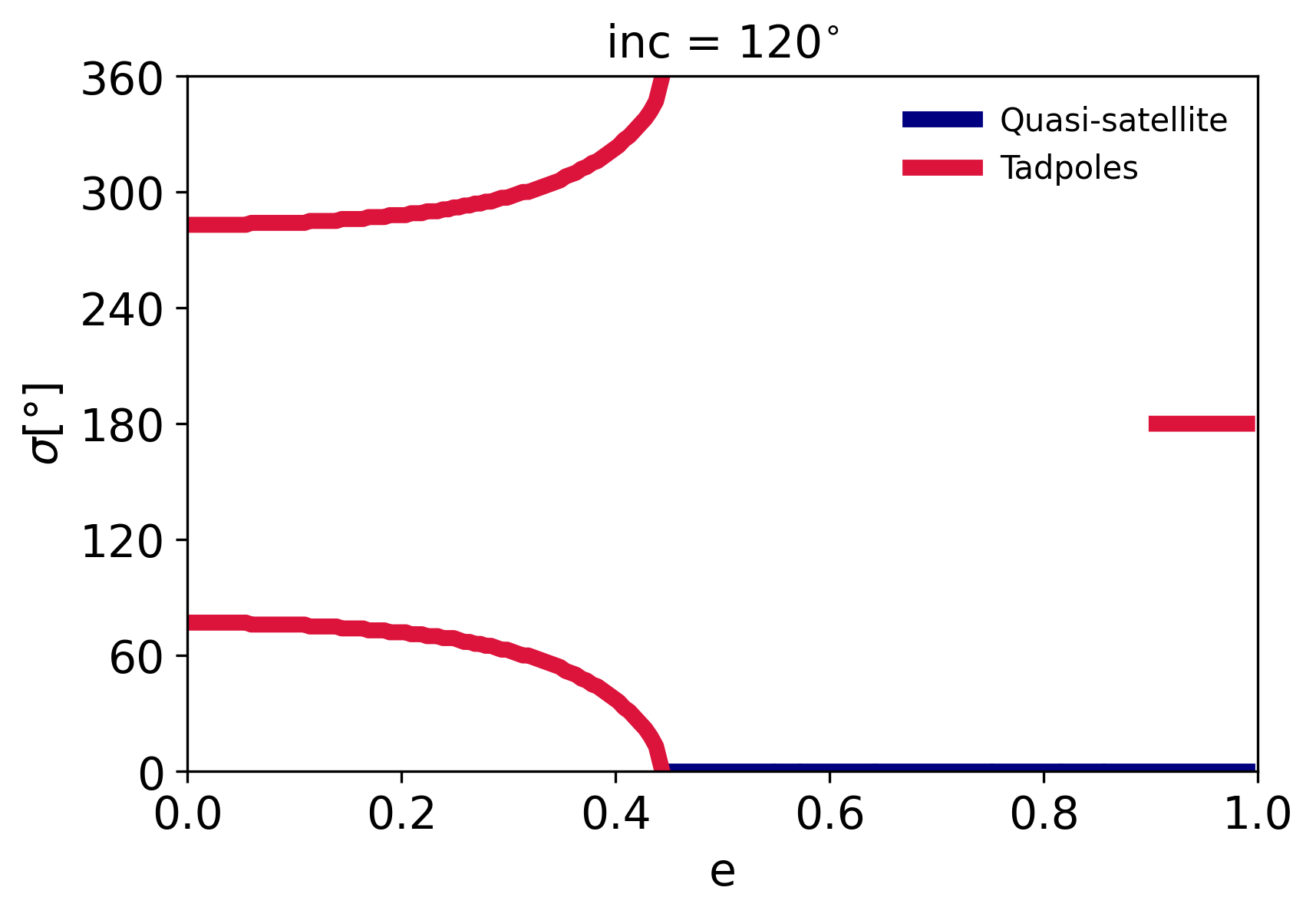} \\
    \includegraphics[width=0.45\linewidth]{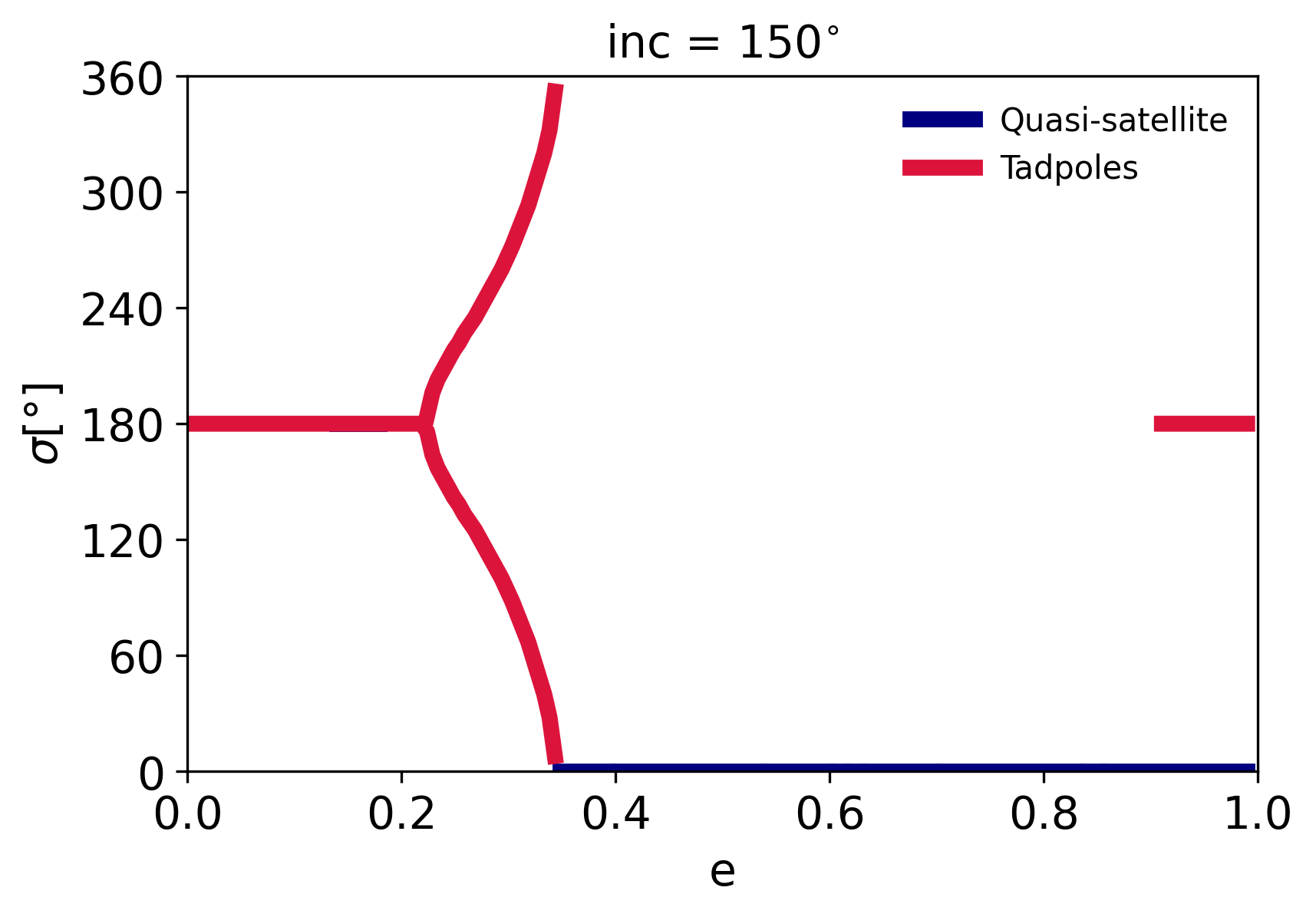} & \includegraphics[width=0.45\linewidth]{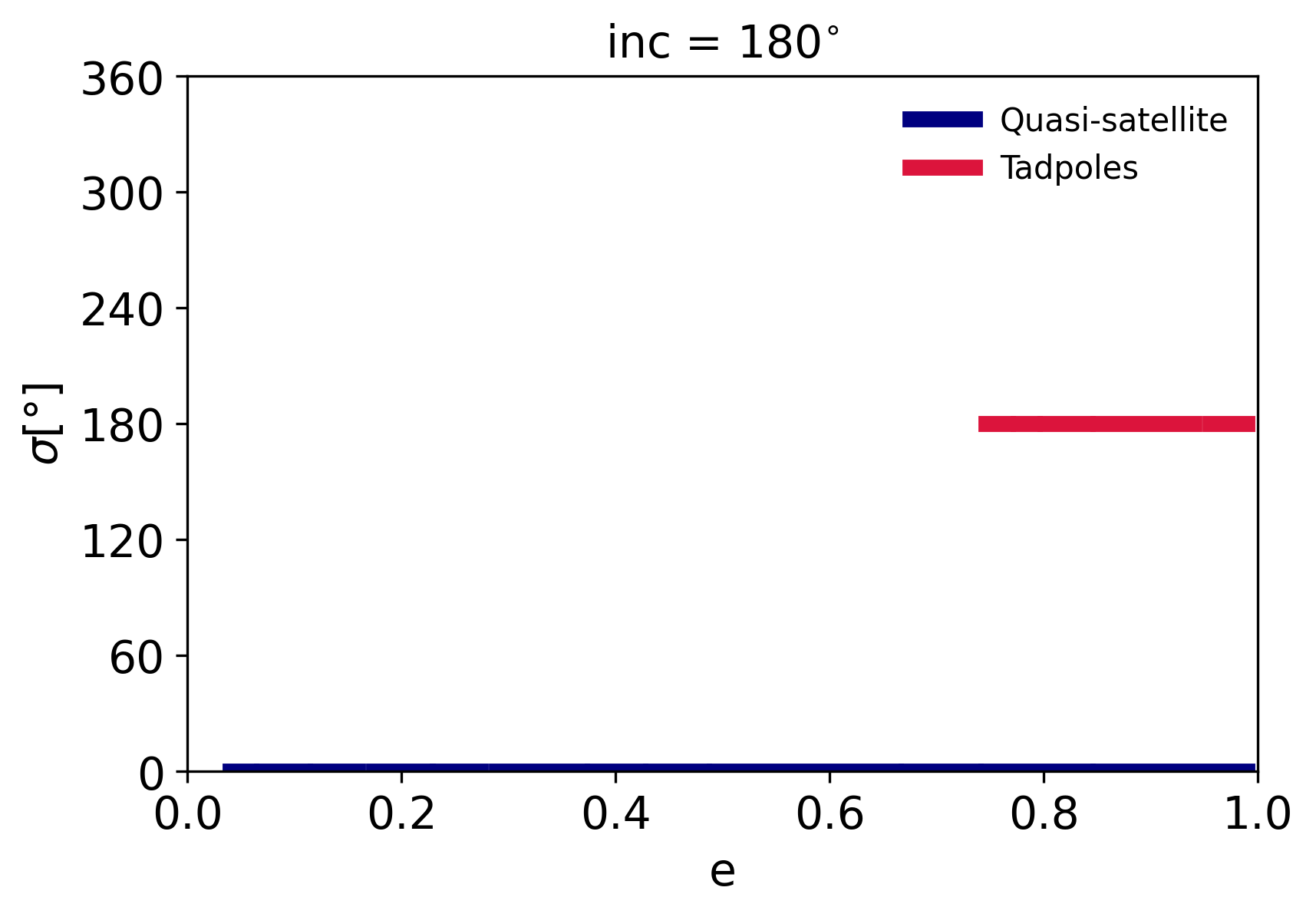} \\
  \end{tabular}
  \caption{Location of stable equilibrium points for the 1:1 resonance with a fictitious planet in circular orbit ($m_{p}=0.001 M_{\odot}$ and $a_{p} = 5.0$ au) as a function of both eccentricity and inclination of a particle assuming $\omega = \Omega = 0^{\circ}$. Polar orbits ($i=90^{\circ}$) are  similar to the $i=120^{\circ}$ case.}
  \label{fig:1_1_sigma}
\end{figure}

The initial application of our model involved characterizing the resonance structure by observing how the equilibrium points shift as the object's orbital elements depart from the circular co-planar case.
Figure \ref{fig:1_1_sigma} illustrates how the equilibrium points' location change for a particle with different values in the full range of values of $(e,i)$, while fixing ($\omega$,$\Omega$) to zero.

For low inclinations and low eccentricities, we successfully recovered the  known L4 and L5 points, with $\sigma \sim 60^{\circ}$ or $\sigma \sim 300^{\circ}$, respectively.
Notably, for higher eccentricities L4 and L5 move depending on the particle's inclination. At low inclinations, the two points progressively converge as the eccentricity increases, eventually merging when the eccentricity is close to 0.9.
On the other hand, for inclinations higher than 30 degrees, the equilibrium points move outward until they vanish, resulting in the no existence  of Trojan-like orbits.
Note that Quasi-satellite solutions located in $\sigma = 0^{\circ}$ still exist even for extreme eccentric and inclined configurations. We have checked the existence of these equilibrium points with numerical integrations of fictitious particles.
Using an analytical approach that expands the resonant disturbing function up to fourth order in ($e,i$), \cite{2000CeMDA..76..131N} demonstrated that the equilibrium points at $\sigma$ = 60$^{\circ}$ and $\sigma$ = 300$^{\circ}$ are shifted in the circular restricted three-body problem for eccentric and inclined particles.
Similar results can be found in \cite{2001Icar..153..391B}, where the authors showed the dependence of the libration center for eccentricities up to 0.3. Another  conclusion that we can draw looking at Figure \ref{fig:1_1_sigma} is that the Tadpoles tend to disappear as the orbital inclination increases.
The Hamiltonian phase space ($\sigma$, $a$) provides a clearer perspective on the changes in equilibrium points.
As an example of this, we consider the upper right panel of Figure \ref{fig:1_1_sigma} that corresponds to $i = 30^{\circ}$ and see how the minima and level curves of the Hamiltonian move, appear and disappear. 
These results are consistent with the phase portraits presented in \citet{2002CeMDA..82..323N}. Although not explicitly shown here, the model used in this work can reproduce changes in the topology of the resonance for different inclinations, arguments of perihelion and planetary eccentricities. This can be reproduced running the codes given in Appendix \ref{app:github}.

\begin{figure}[H]
  \centering
  \begin{tabular}{cc}
    \includegraphics[width=0.45\linewidth]{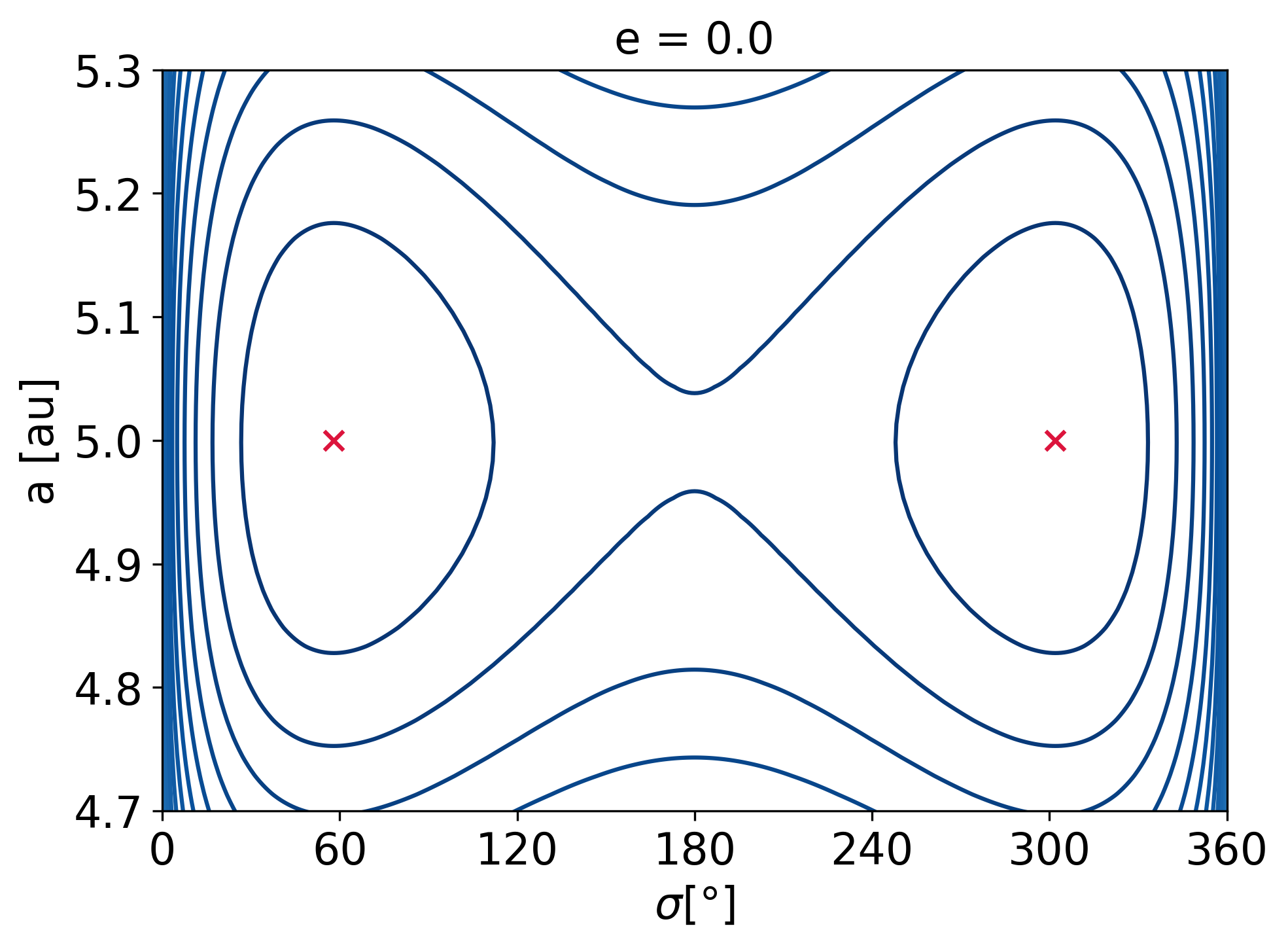} & \includegraphics[width=0.45\linewidth]{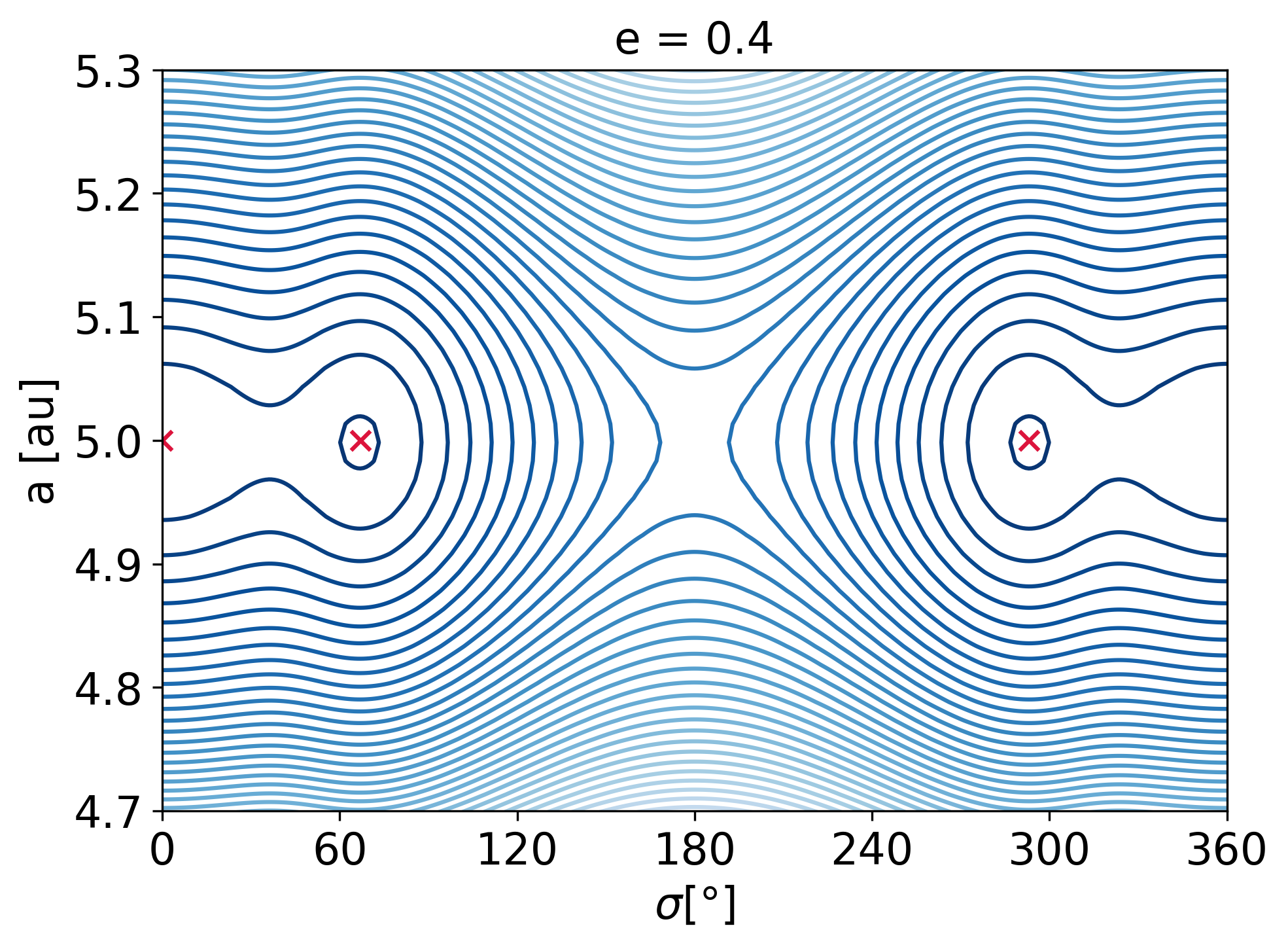} \\
    \includegraphics[width=0.45\linewidth]{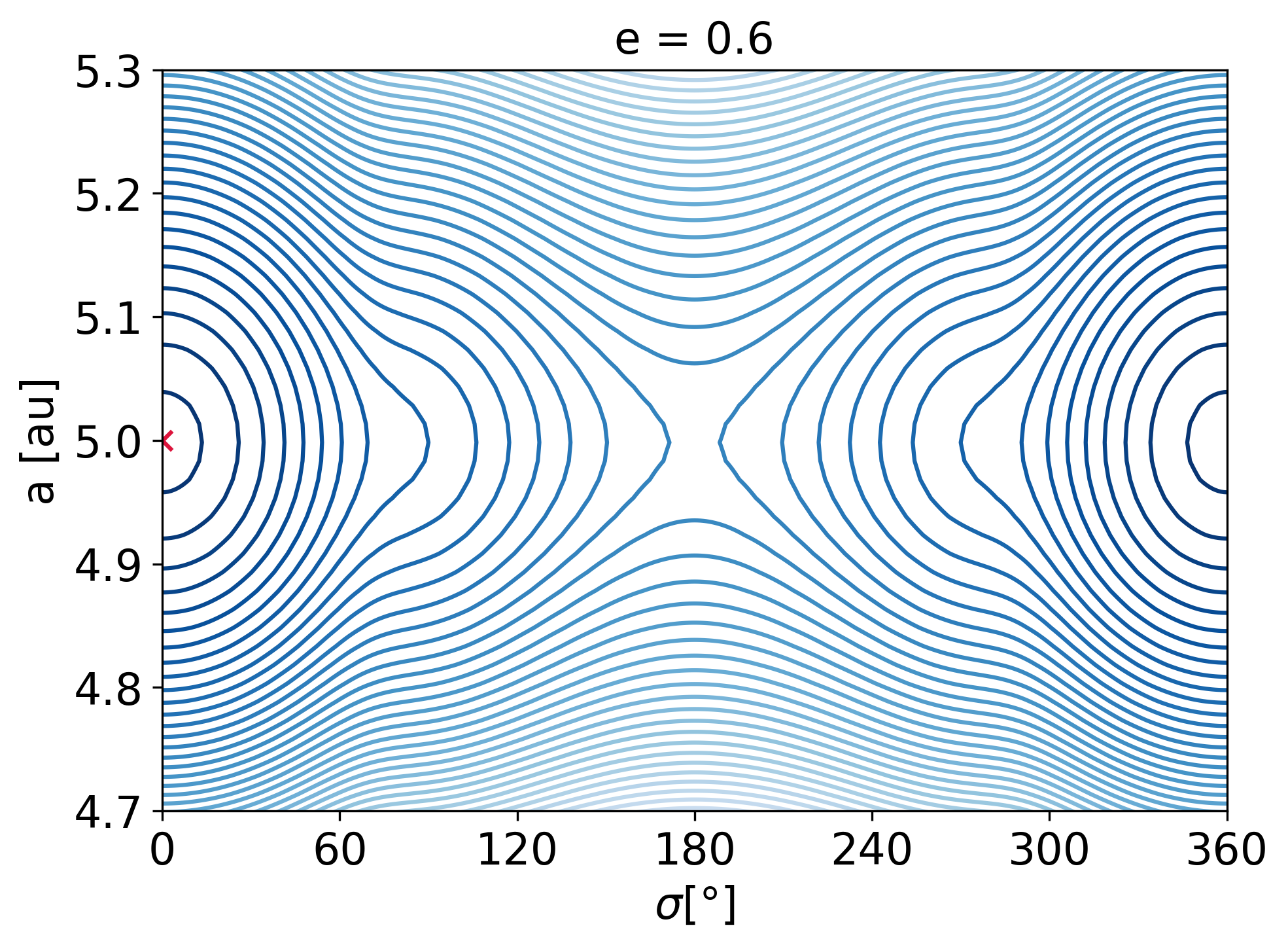} & \includegraphics[width=0.45\linewidth]{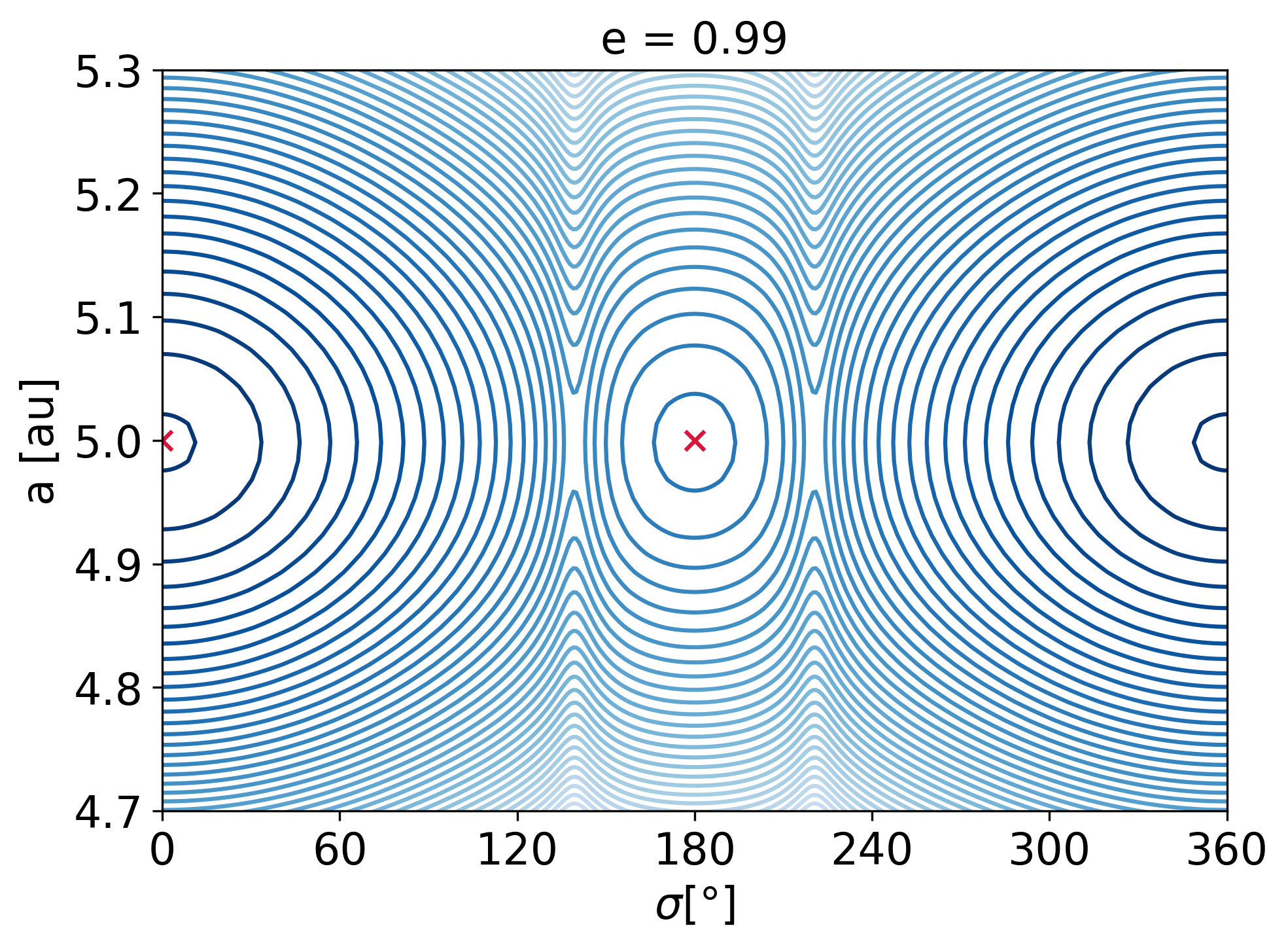} \\
  \end{tabular}
  \caption{Phase space Hamiltonian level curves ($\sigma ,a$) for an asteroid of orbital elements $i = 30^{\circ}, \omega = 0^{\circ}$ and varying eccentricities. All level curves represent the 1:1 resonance with a circular planet of $m_{p}=0.001 M_{\odot}$ and $a_{p} = 5.0$ au. Stable equilibrium points are denoted by red crosses.}
  \label{fig:ham_points_change}
\end{figure}

This Hamiltonian portrait helps us understand why some stable equilibrium points disappear. For the case of $i = 30^{\circ}$, we observe that the two stable equilibrium points vanish at an eccentricity value around $e \sim 0.47$.
This is intriguing, as equilibrium points can only disappear if they collide with unstable equilibrium points. Indeed, this is what happens in this scenario.
In the second panel of Figure \ref{fig:ham_points_change}, for an eccentricity of $e = 0.4$, the system has three stable and three unstable equilibrium points. However, in the third panel (corresponding to $e = 0.6$), the system only has one stable and one unstable equilibrium point. The disappearance of some equilibrium points within this range of eccentricities results from their collision with the unstable ones. We have included an animation as complementary material in order to make the situation clearer.

\begin{figure}[H]
    \centering
    \includegraphics[scale=0.60]{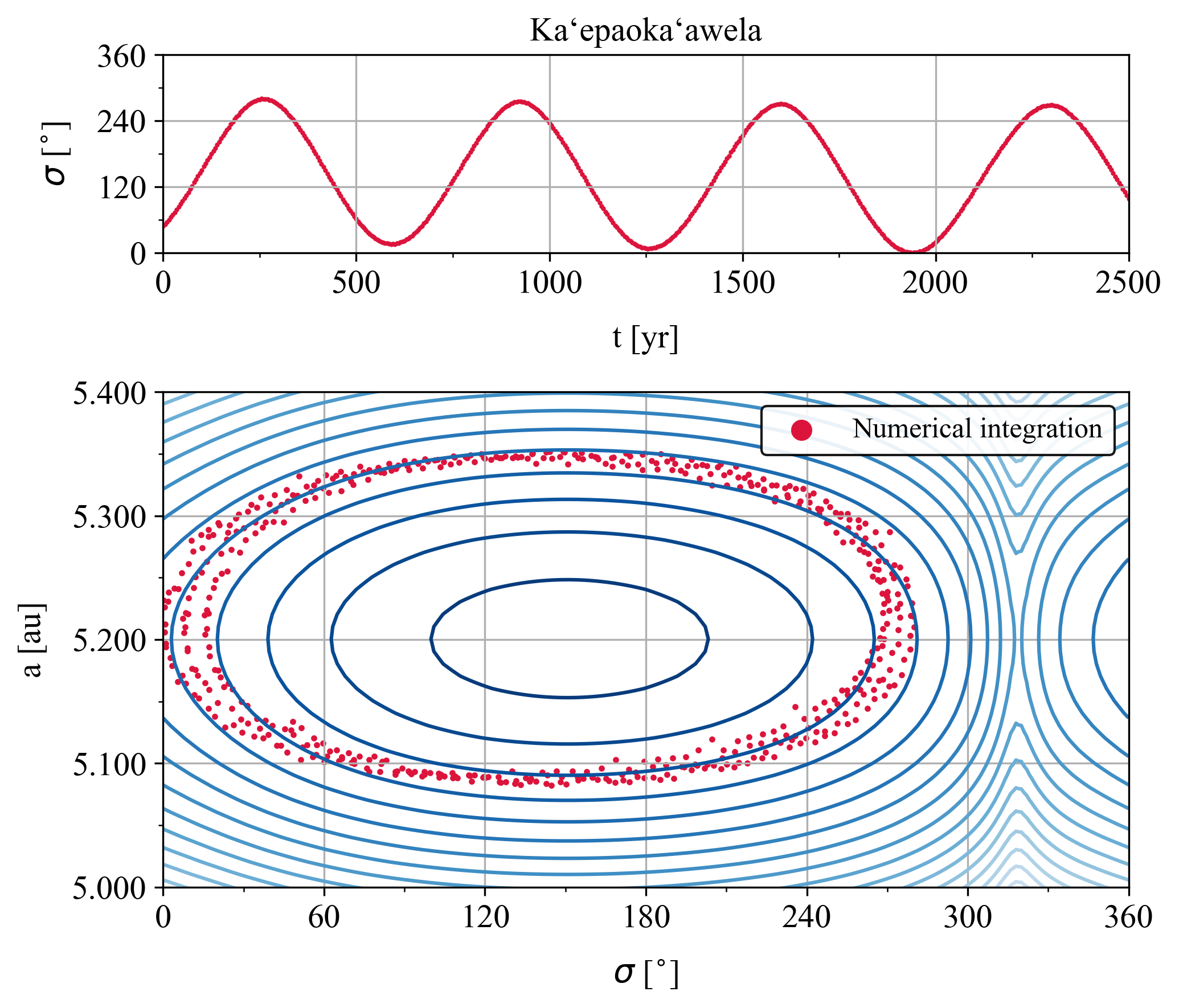}
    \caption{Top: N-body integration showing the critical angle of Ka\text{\textquoteleft}epaoka\text{\textquoteleft}awela (with $e$ $\sim$ 0.38 and $i$ $\sim$ 163$^{\circ}$) over 2500 years. Bottom: same numerical integration (Red) plotted with the Hamiltonian level curves (Blue) indicating that the resonant angle librates around the predicted value given by the theory.}
    \label{fig:kaepa}
\end{figure}

The location of equilibrium points is an important factor to be taken into account. As discussed above, these points can vary significantly depending on the resonant configuration of the system.
As an extreme example, we present in Figure \ref{fig:kaepa} the time evolution of the first confirmed retrograde Jupiter co-orbital, Ka\text{\textquoteleft}epaoka\text{\textquoteleft}awela \citep{2017Natur.543..687W}. While Jupiter's co-orbitals are not included in the following analysis, we highlight this case to underscore the significance of retrograde co-orbital objects, which, to date, have only been discovered for Jupiter.
It is noteworthy that we can study time evolution of any orbit regardless of the high inclination or eccentricity. In this case, this object has $e \sim 0.38$ and $i \sim 163.0^{\circ}$.

In Figure \ref{fig:kaepa}, color red represents the numerical solution for the evolution of the retrograde co-orbital under the perturbations of the real planetary system on top of the numerically calculated Hamiltonian.
According to the semi-analytic theory considering only Jupiter with its real eccentricity, the predicted small amplitude libration period for $\sigma$ is $787$ years. However, the numerically integrated particle librates with a period of $685$ years.
This discrepancy arises from the fact that the libration does not occur precisely at the center, resulting in oscillations with a large amplitude that decrease the period.
The model's Hamiltonian approximately describes the evolution with enough accuracy for our present purpose, despite perturbations from other planets. As changes in the orbital elements of the particle also change the level curves, this needs to be taken into account in order to track the trajectory over extended time scales. 
The secular evolution of the particle can be followed but not predicted by this theory. An example is included as complementary material.

After the numerical integrations described in Section \ref{sec:methods} were performed, we visually determined which objects have a resonant behavior with librating resonant angles and a path on top of a resonant Hamiltonian level curve.
Tables \ref{tab:test_table} $\&$ \ref{tab:test_table_2} presents a list of objects identified to have this type of orbit with a planet in the inner Solar System, whereas Table \ref{tab:test_table_3} lists those resonating with outer Solar System planets, excluding Jupiter due the large number of co-orbitals.
We have also included an orbit type classification based on the evolution of the N body integrations over the Hamiltonian level curves and the condition code of the orbit in order to see which determinations may be fictitious due to lack of observations.
Due to this high uncertainties in some asteroid orbits, statistical analysis using clones is necessary to faithfully determine whether an object is in resonance with a planet.
This approach is crucial until more precise observations are available to further constrain the orbital elements.
A comprehensive study of each real orbit is beyond the scope of this work, as other effects such as Yarkovsky were not considered. Therefore, we present only an initial attempt to classify all Solar System co-orbital orbits integrating the osculating orbital elements.

\newpage

\begin{table}[h!]
    \begin{subtable}{0.30\textwidth}
    \renewcommand{\arraystretch}{1.3}
    \centering
    \begin{tabular}{|c|c|c|c|}
        \hline
        Object & Planet & Orbit & CC \\
        \hline
        \hline
        2001 CK$_{32}$ & Venus & HS-QS & 0 \\
        Zoozve & Venus  & QS & 0  \\
        2012 XE$_{133}$ & Venus  & L5 & 0  \\
        2013 ND$_{15}$ & Venus & L4 & 7 \\
        2015 WZ$_{12}$ & Venus & jumping & 6 \\
        2019 HH$_{3}$ & Venus & sticking & 7 \\
        2020 CL$_{1}$ & Venus  & QS & 1 \\
        2020 SB & Venus  & HS-QS & 1 \\
        2020 SV$_{5}$ & Venus & jumping & 7 \\
        2022 BL$_{5}$ & Venus  & QS & 7 \\
        2022 CD & Venus  & jumping  & 5 \\
        2022 UA$_{13}$ & Venus & jumping & 6 \\
        2023 BB$_{1}$ & Venus  & jumping & 7 \\
        2023 QS$_{7}$ & Venus & jumping & 8 \\
        2024 AF$_{6}$ & Venus & HS-QS & 6 \\
        \hline
        Cruithne & Earth  & HS-QS & 0 \\
        2001 GO$_{2}$ & Earth & HS-QS & 7 \\
        2002 AA$_{29}$ & Earth & HS & 0 \\
        2004 GU$_{9}$ & Earth & QS & 0 \\
        2005 QQ$_{87}$ & Earth & jumping & 0  \\
        2006 FV$_{35}$ & Earth & QS & 1 \\
        2008 WM$_{64}$ & Earth & sticking & 0 \\
        2009 HE$_{60}$ & Earth & HS & 7 \\
        2010 NY$_{65}$ & Earth & sticking & 0 \\
        2010 SO$_{16}$ & Earth & HS & 0 \\
        2010 TK$_{7}$ & Earth & L4 & 0 \\
        2013 LX$_{28}$ & Earth & QS & 1 \\
        2014 EK$_{24}$ & Earth & sticking & 0 \\
        2014 OL$_{339}$ & Earth & HS-QS & 1 \\
        2015 SO$_{2}$ & Earth & HS-QS & 1 \\
        2015 XX$_{169}$ & Earth & HS-QS & 0 \\
        2015 YA & Earth & QS & 6 \\
        2016 CA$_{138}$ & Earth & HS-QS & 0 \\
        2016 CO$_{246}$ & Earth & HS-QS & 0 \\
        \hline
    \end{tabular}
    \end{subtable}
    \hspace{5mm}
    \begin{subtable}{0.30\textwidth}
    \renewcommand{\arraystretch}{1.3}
    \centering
    \begin{tabular}{|c|c|c|c|}
        \hline
        Object & Planet & Orbit & CC \\
        \hline
        \hline
        2016 FU$_{12}$ & Earth & sticking & 7 \\
        Kamo\text{\textquoteleft}oalewa & Earth  & HS-QS & 0 \\
        2016 JP & Earth & jumping & 0 \\
        2017 SL$_{16}$ & Earth & HS-QS & 0 \\
        2017 XQ$_{60}$ & Earth & HS-QS & 0 \\
        2018 AN$_{2}$ & Earth & HS-QS & 7 \\
        2018 XW$_{2}$ & Earth & HS-QS & 7 \\
        2019 GM$_{1}$ & Earth & HS-QS & 0 \\
        2019 HS$_{2}$ & Earth & sticking & 7 \\
        2019 NC$_{1}$ & Earth & sticking & 0 \\
        2019 SB$_{6}$ & Earth & HS-QS & 7 \\
        2019 VL$_{5}$ & Earth & HS & 0 \\
        2019 XH$_{2}$ & Earth & HS-QS & 6 \\
        2019 YB$_{4}$ & Earth & HS & 6 \\
        2020 CX$_{1}$ & Earth & HS-QS & 0 \\
        2020 DX$_{1}$ & Earth & sticking & 7 \\
        2020 HE$_{5}$ & Earth & sticking & 5 \\
        2020 PN$_{1}$ & Earth & HS-QS & 0 \\
        2020 PP$_{1}$ & Earth & HS-QS & 0 \\
        2020 XL$_{5}$ & Earth & L4 & 0 \\
        2021 BA & Earth & HS-QS & 1 \\
        2021 GN$_{1}$ & Earth & sticking & 8 \\
        2021 VU$_{12}$ & Earth & sticking & 1 \\
        2021 XS$_{4}$ & Earth & sticking & 7 \\
        2022 UO$_{10}$ & Earth & sticking & 7 \\
        2022 UY & Earth & sticking & 1 \\
        2022 UP$_{20}$ & Earth & sticking & 2 \\
        2022 VR$_{1}$ & Earth & HS-QS & 0 \\
        2022 YG & Earth & HS-QS & 2 \\
        2023 FW$_{13}$ & Earth & QS & 0 \\
        2023 GC$_{2}$ & Earth & HS-QS & 7 \\
        2023 QR$_{1}$ & Earth & HS-QS & 6 \\
        2023 TG$_{14}$ & Earth & HS-QS & 0 \\
        2024 AV$_{2}$ & Earth & jumping & 1 \\
        2024 JR$_{16}$ & Earth & jumping & 7 \\
        \hline
    \end{tabular}
    \end{subtable}
    \caption{Asteroids identified in 1:1 resonance with Venus and Earth. Orbit type is based on the evolution of the critical angle over the Hamiltonian level curves. Jumping objects are more unstable. See text for the class ``sticking''.}
    \label{tab:test_table}
\end{table}

\newpage

\begin{table}[h!]
    \begin{subtable}{0.30\textwidth}
    \renewcommand{\arraystretch}{1.3}
    \centering
    \begin{tabular}{|c|c|c|c|}
        \hline
        Object & Planet & Orbit & CC \\
        \hline
        \hline
        Eureka & Mars & L5 & 0 \\
        1998 VF$_{31}$ & Mars  & L5 & 0 \\
        1999 ND$_{43}$ & Mars & jumping & 0 \\
        1999 UJ$_{7}$ & Mars  & L4 & 0 \\
        2000 XH$_{47}$ & Mars & sticking & 0 \\
        2001 DH$_{47}$ & Mars & L5 & 0 \\
        2001 FG$_{24}$ & Mars & sticking & 0 \\
        2006 XY$_{2}$ & Mars & HS-QS & 6 \\
        2007 NS$_{2}$ & Mars & L5 & 0 \\
        2007 UR$_{2}$ & Mars & HS & 0 \\
        2009 SE & Mars & L5 & 3 \\
        2010 RL$_{82}$ & Mars & sticking & 0 \\
        2010 XH$_{11}$ & Mars & sticking & 6 \\
        2011 SC$_{191}$ & Mars & L5 & 0 \\
        2011 SL$_{25}$ & Mars & L5 & 1 \\
        2011 SP$_{189}$ & Mars & L5 & 2 \\
        2011 UB$_{256}$ & Mars & L5 & 0 \\
        2011 UN$_{63}$ & Mars & L5 & 0 \\
        2012 QR$_{50}$ & Mars & sticking & 1 \\
        2013 PM$_{43}$ & Mars & sticking & 0 \\
        2014 HM$_{187}$ & Mars & HS-QS & 0 \\
        2014 JU$_{24}$ & Mars & jumping & 6 \\
        2014 SA$_{224}$ & Mars & jumping & 6 \\
        2015 CW$_{12}$ & Mars & jumping & 6 \\
        \hline
    \end{tabular}
    \end{subtable}
    \hspace{5mm}
    \begin{subtable}{0.30\textwidth}
    \renewcommand{\arraystretch}{1.3}
    \centering
    \begin{tabular}{|c|c|c|c|}
        \hline
        Object & Planet & Orbit & CC \\
        \hline
        \hline
        2015 TL$_{144}$ & Mars & L5 & 2 \\
        2016 AA$_{165}$ & Mars & L5 & 0 \\
        2016 CP$_{31}$ & Mars & L5 & 1 \\
        2016 NZ$_{55}$ & Mars & L4 & 2 \\
        2016 QY$_{10}$ & Mars & L4 & 5 \\
        2017 QW$_{35}$ & Mars & HS & 3 \\
        2017 XG$_{62}$ & Mars & QS & 0 \\
        2018 EC$_{4}$ & Mars & L5 & 1 \\
        2018 FC$_{4}$ & Mars & L5 & 3 \\
        2018 FM$_{29}$ & Mars & L5 & 1 \\
        2019 BG$_{3}$ & Mars & sticking & 6 \\
        2019 KF$_{1}$ & Mars & HS & 2 \\
        2020 JO$_{1}$ & Mars & jumping & 7 \\
        2020 LE$_{1}$ & Mars & QS & 0 \\
        2020 VT$_{1}$ & Mars & HS-QS & 6 \\
        2021 FV$_{1}$ & Mars & QS & 8 \\
        2021 JK$_{4}$ & Mars & HS-QS & 1 \\
        2021 TB$_{3}$ & Mars & HS-QS & 7 \\
        2021 WX$_{6}$ & Mars & jumping & 2 \\
        2022 OG$_{2}$ & Mars & HS & 5 \\
        2023 FW$_{14}$ & Mars & L4 & 0 \\
        2023 QS$_{3}$ & Mars & HS & 7 \\
        2023 RJ$_{6}$ & Mars & HS-QS & 7 \\
        2023 WV$_{1}$ & Mars & jumping & 8 \\
        \hline
    \end{tabular}
    \end{subtable}
    \caption{Asteroids identified in 1:1 resonance with Mars. Orbit type is based on the evolution of the critical angle over the Hamiltonian level curves. Jumping objects are more unstable. See text for the class ``sticking''.}
    \label{tab:test_table_2}
\end{table}

\newpage

\begin{table}[h!]
    \begin{subtable}{0.30\textwidth}
    \renewcommand{\arraystretch}{1.3}
    \centering
    \begin{tabular}{|c|c|c|c|}
        \hline
        Object & Planet & Orbit & CC \\
        \hline
        \hline
        1999 RG$_{33}$ & Saturn & QS & 1 \\
        2001 BL$_{41}$ & Saturn & sticking & 0 \\
        2017 SV$_{13}$ & Saturn & jumping & 6 \\
        2019 UO$_{14}$ & Saturn & L4 & 2 \\
        \hline
        Crantor & Uranus & HS-QS & 1 \\
        2002 VG$_{131}$ & Uranus & QS & 9 \\
        2011 QF$_{99}$ & Uranus & L4 & 2 \\
        2012 DS$_{85}$ & Uranus & HS-QS & 5 \\
        2013 UC$_{17}$ & Uranus & sticking & 3 \\
        2014 YX$_{49}$ & Uranus & L4 & 1 \\
        2022 OH$_{10}$ & Uranus & HS-QS & 2 \\
        2023 UV$_{10}$ & Uranus & jumping & 2 \\
        \hline
        2001 QR$_{322}$ & Neptune & L4 & 3 \\
        2004 KV$_{18}$ & Neptune & L5 & 4 \\
        Otrera & Neptune & L4 & 4 \\
        2005 TN$_{53}$ & Neptune & L4 & 4 \\
        Clete & Neptune & L4 & 4 \\
        2006 RJ$_{103}$ & Neptune & L4 & 2 \\
        2007 RW$_{10}$ & Neptune & QS & 2 \\
        2007 VL$_{305}$ & Neptune & L4 & 2 \\
        2008 LC$_{18}$ & Neptune & L5 & 4 \\
        2010 DF$_{106}$ & Neptune & sticking & 2 \\
        2010 EN$_{65}$ & Neptune & jumping & 2 \\
        2010 TS$_{191}$ & Neptune & L4 & 2 \\
        2010 TT$_{191}$ & Neptune & L4 & 4 \\
        \hline
    \end{tabular}
    \end{subtable}
    \hspace{5mm}
    \begin{subtable}{0.30\textwidth}
    \renewcommand{\arraystretch}{1.3}
    \centering
    \begin{tabular}{|c|c|c|c|}
        \hline
        Object & Planet & Orbit & CC \\
        \hline
        \hline
        2011 HM$_{102}$ & Neptune & L5 & 3 \\
        2011 SO$_{277}$ & Neptune & L4 & 3 \\
        2011 WG$_{157}$ & Neptune & L4 & 3 \\ 
        2012 UD$_{185}$ & Neptune & L4 & 2 \\ 
        2012 UV$_{177}$ & Neptune & L4 & 3 \\ 
        2012 UW$_{177}$ & Neptune & L4 & 5 \\ 
        2013 KY$_{18}$ & Neptune & L5 & 2 \\
        2013 RC$_{158}$ & Neptune & L4 & 4 \\ 
        2013 RL$_{124}$ & Neptune & L4 & 4 \\ 
        2013 TK$_{227}$ & Neptune & L4 & 4 \\ 
        2013 TZ$_{187}$ & Neptune & L4 & 3 \\ 
        2013 VX$_{30}$ & Neptune & L4 & 3 \\ 
        2014 QO$_{441}$ & Neptune & L4 & 4 \\ 
        2014 QP$_{441}$ & Neptune & L4 & 4 \\ 
        2014 RO$_{74}$ & Neptune & L4 & 3 \\ 
        2014 SC$_{374}$ & Neptune & L4 & 4 \\ 
        2014 UU$_{240}$ & Neptune & L4 & 4 \\ 
        2014 YB$_{92}$ & Neptune & L4 & 4 \\ 
        2015 RW$_{277}$ & Neptune & L4 & 5 \\ 
        2015 VU$_{207}$ & Neptune & L4 & 3 \\ 
        2015 VV$_{165}$ & Neptune & L4 & 3 \\ 
        2015 VW$_{165}$ & Neptune & L4 & 3 \\ 
        2015 VX$_{165}$ & Neptune & L4 & 4 \\ 
        2019 GA$_{143}$ & Neptune & L5 & 9 \\ 
        2022 LP$_{15}$ & Neptune & HS & 9 \\ 
        \hline
    \end{tabular}
    \end{subtable}
    \caption{Asteroids identified in 1:1 resonance with outer Solar System planets. Orbit type is based on the evolution of the critical angle over the Hamiltonian level curves. Jumping objects are more unstable. See text for the class ``sticking''.}
    \label{tab:test_table_3}
\end{table}


We successfully identified all previously confirmed co-orbitals and additionally confirmed new ones, categorizing them according to their specific type of co-orbital orbit.
The number of identified objects in co-orbital orbits, excluding Jupiter, amounts to 117 for the inner Solar System and 50 for the outer Solar System, resulting in a combined total of 167.
We were able to classify some objects in an orbit class labeled sticking, characterized by the resonant angle circulating but the orbital evolution being significantly influenced by the resonance.
The sticking evolution is a known phenomena that has been observed in several resonances \citep{2006P&SS...54...87L, 2007Icar..192..238L}.
It is characterized by the evolution to one side of the resonance, showing some stability against other perturbations in despite not being strictly captured in the resonance.
\citet{2012AcA....62..113W} reported two Jupiter co-orbitals 2006 UG$_{185}$ and 2000 HR$_{24}$ that show short-lived sticking orbits.
As an arbitrary requierment for including a sticking object in the presented list, we only account for those whose semi-major axis change is at least one-fifth of the resonance width.
Detailed analyses of each orbit type, along with an overview, are presented in the next sections.
We provide detailed descriptions of several newly confirmed objects as well as other interesting objects organized by orbit type.

\subsection{Objects in Tadpole orbits}

Arguably the most known co-orbital orbit, objects in Tadpole orbits librate around asymmetric equilibrium points of the Hamiltonian. From our dynamical classification, we found that all Solar System planets have Tadpole companions with the exception of Mercury.

\begin{figure}[H]
    \centering
    \begin{subfigure}{0.4\textwidth}
        \centering
        \includegraphics[scale=0.42]{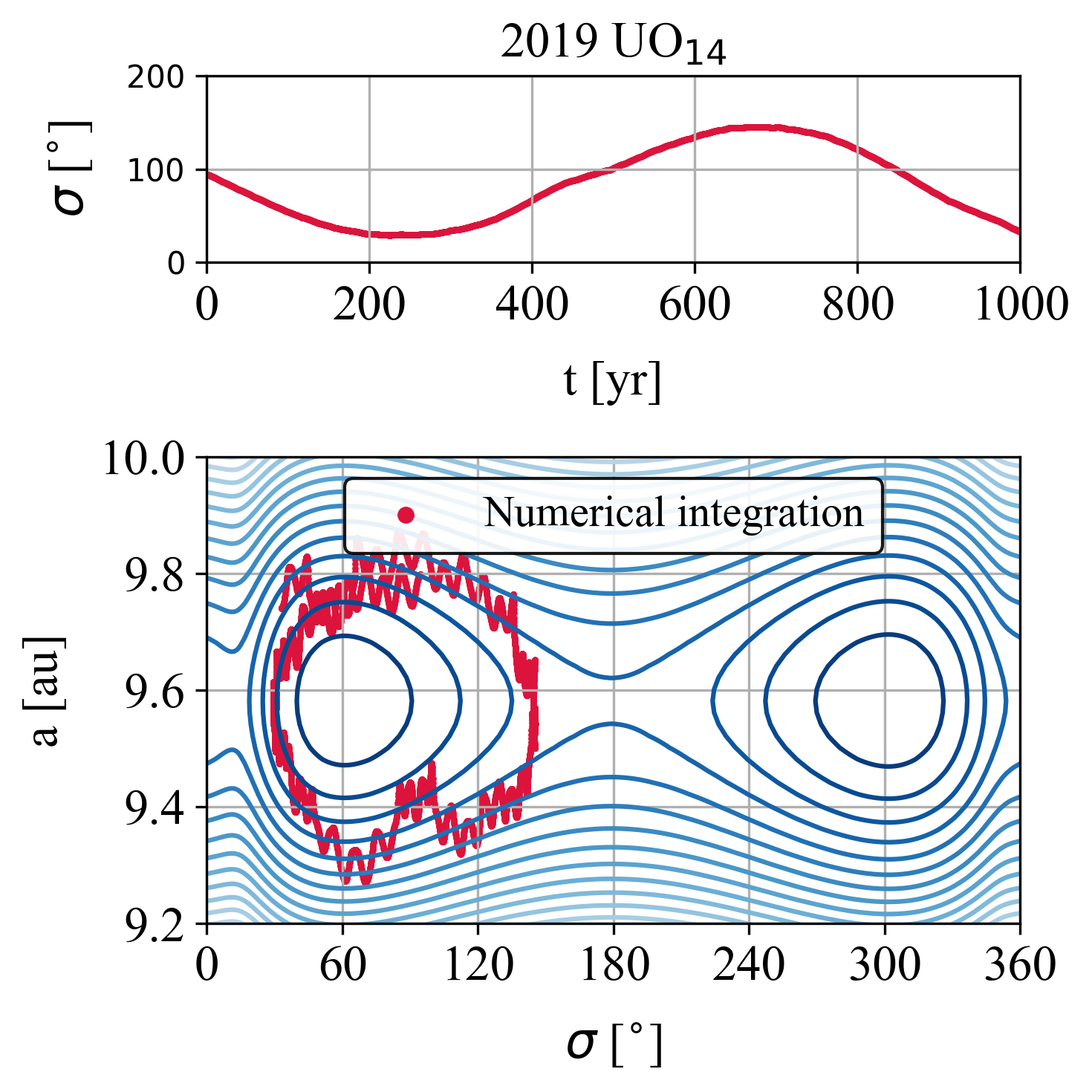}
        \caption{}
        \label{fig:plot_tadpole_a}
    \end{subfigure}
    \begin{subfigure}{0.4\textwidth}
        \centering
        \includegraphics[scale=0.42]{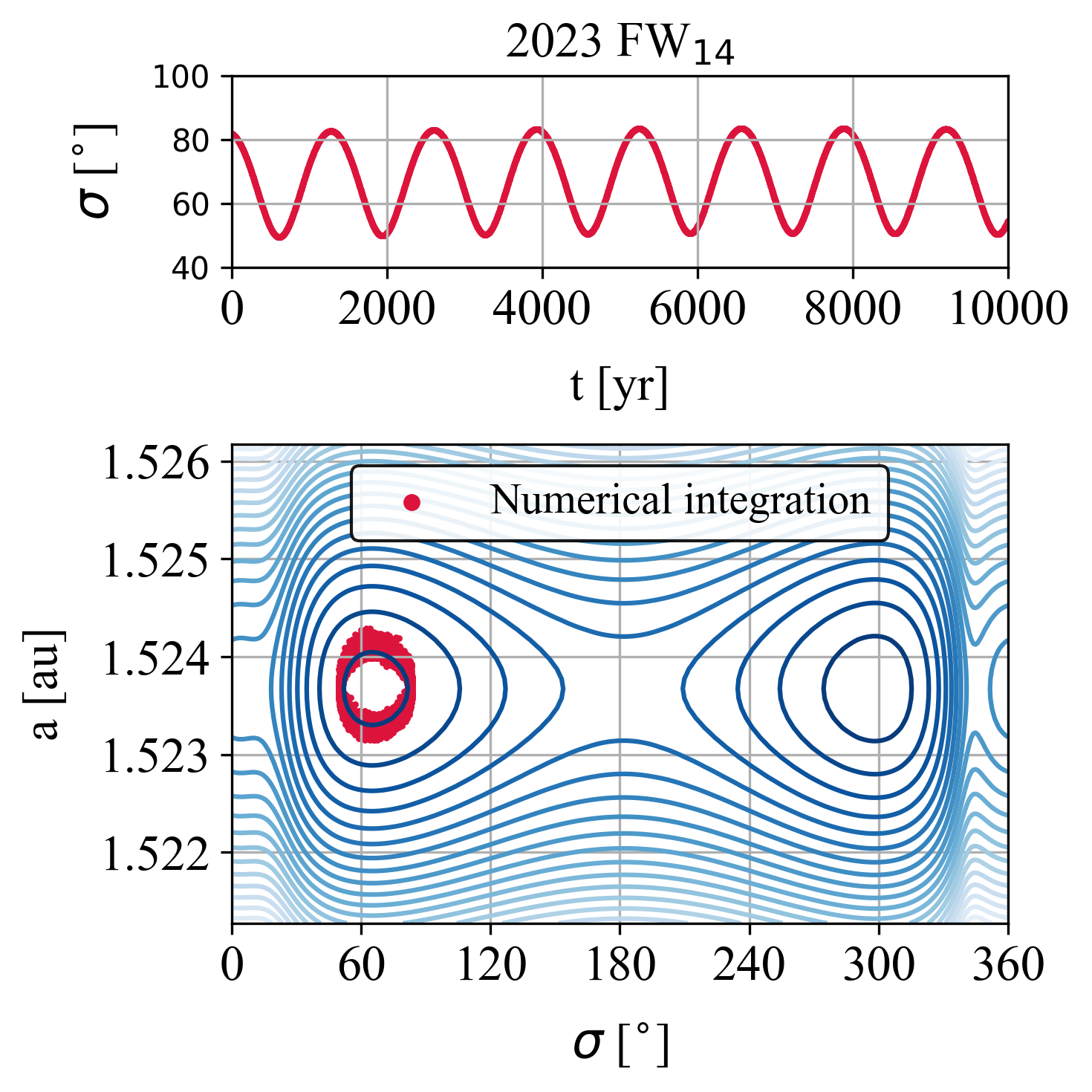}
        \caption{}
        \label{fig:plot_tadpole_b}
    \end{subfigure}
    \caption{(a) Time evolution of 2019 UO$_{14}$ ($e \sim 0.24 $, $i \sim 33^{\circ}$). This object will remain in a Tadpole orbit with Saturn for only one libration period. (b) Time evolution of asteroid 2023 FW$_{14}$ ($e = 0.16$, $i = 13^{\circ}$), a recently discovered object in a Tadpole orbit with Mars.}
    \label{fig:plot_tadpole}
\end{figure}

Despite the fact that Saturn is the second most massive planet in the Solar System we currently do not have a known co-orbital population as large as that of Jupiter.
This is primarily due to secular resonances and Jupiter's perturbations, which destabilize objects in Saturn's co-orbital regions \citep{2014MNRAS.437.1420H}.
Consequently, all Saturn co-orbitals are highly transient.
As an example of this, we present in Figure \ref{fig:plot_tadpole_a} the orbit evolution of 2019 UO$_{14}$, which we discovered to be a Saturn Trojan located at the Lagrange point L4. During the peer review process of this work, \citet{2024arXiv240919725H} reached similar conclusions regarding this object.

So far, 1999 UJ$_{7}$ was the only know Mars L4 Trojan. However, we discovered that 2023 FW$_{14}$ is the second object to be in such orbit.
Figure \ref{fig:plot_tadpole} shows the orbital evolution of the new Saturn and Mars Trojans.

The trayectory of 2023 FW$_{14}$ in phase space follows the predicted path. The duration an object remains at one equilibrium points is highly influenced by perturbations from other planets.
2023 FW$_{14}$ appears to be stable for at least 10 kyr as a Mars L4 Trojan.
During the writing of this work, \cite{2024arXiv240304061D} provided a detailed study of this object.
Contrariwise, the Saturn L4 Trojan 2019 UO$_{14}$, despite librating around the L4 equilibrium point, experiences perturbations and drifts away from the resonance only 1kyr after the start of the integration.

\subsection{Objects in Quasi-satellite orbits}

\begin{figure}[H]
    \centering
    \begin{subfigure}{0.4\textwidth}
        \centering
        \includegraphics[scale=0.42]{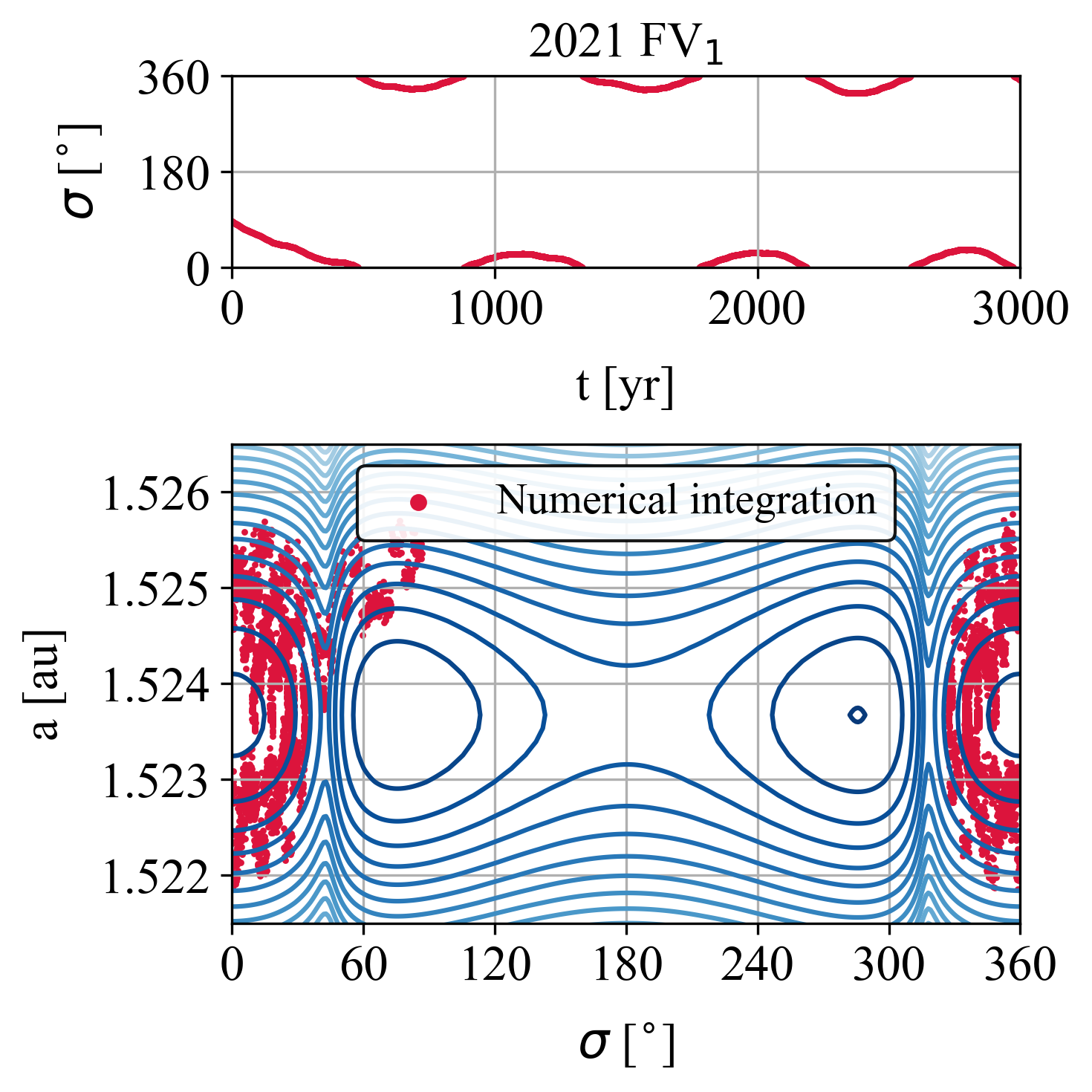}
        \caption{}
        \label{fig:qs_a}
    \end{subfigure}
    \begin{subfigure}{0.4\textwidth}
        \centering
        \includegraphics[scale=0.42]{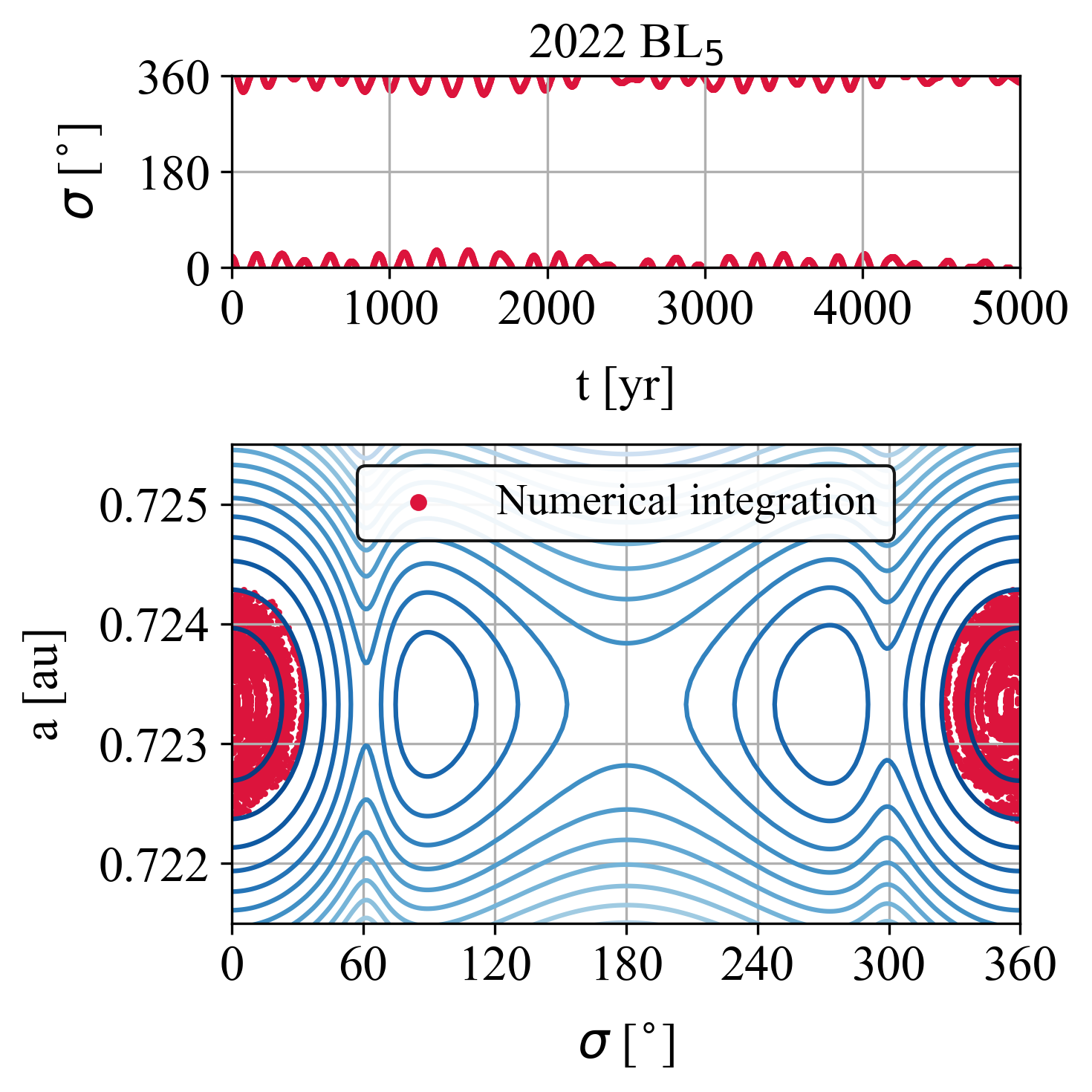}
        \caption{}
        \label{fig:qs_b}
    \end{subfigure}
    \caption{(a) Time evolution of object 2021 FV$_{1}$ ($e \sim  0.28$, $i \sim 4^{\circ}$) as an example of a Quasi-satellite orbit with Mars. (b) Time evolution of object 2022 BL$_{5}$ ($e \sim  0.53$, $i \sim 12^{\circ}$), an example of a Quasi-satellite orbit with Venus.}
    \label{fig:qs}
\end{figure}

Librating around $\sigma = 0^{\circ}$, Quasi-satellite orbits are quite interesting due to their appearance in the frame of reference rotating with the host planet.
We identified that object 2021 FV$_{1}$ is currently in a Quasi-satellite orbit with Mars. As we present in Figure \ref{fig:qs_a}, its critical angle librates around $\sigma \sim 0^{\circ}$.
Numerical integrations indicate that this object will remain in a stable co-orbital trayectory for at least 3 kyr.
Another new Quasi-satellite orbit is the corresponding to 2022 BL$_{5}$, in this case with Venus. As we can see in Figure \ref{fig:qs_b}, the evolution in the phase space librates around $\sigma = 0^{\circ}$
It will remain in such orbit for $\sim$ 26 kyr.

It is noteworthy that both objects show dynamical behavior consistent with the Hamiltonian level curves predicted by the semi-analytical model.

\subsection{Objects in Horseshoe orbits}

Aside from Tadpole and Quasi-satellite orbits, many asteroids have been found to have Horseshoe orbits with several Solar System planets.
As an example of this type of orbit, we present the time evolution of 2022 OG$_{2}$. This object will remain in a Horseshoe orbit for at least 50 kyr.
Another new Horseshoe orbit with Neptune is the one followed by 2022 LP$_{15}$.
It will remain in such orbit for $\sim$ 10 kyr.

\begin{figure}[H]
    \centering
    \begin{subfigure}{0.4\textwidth}
        \centering
        \includegraphics[scale=0.42]{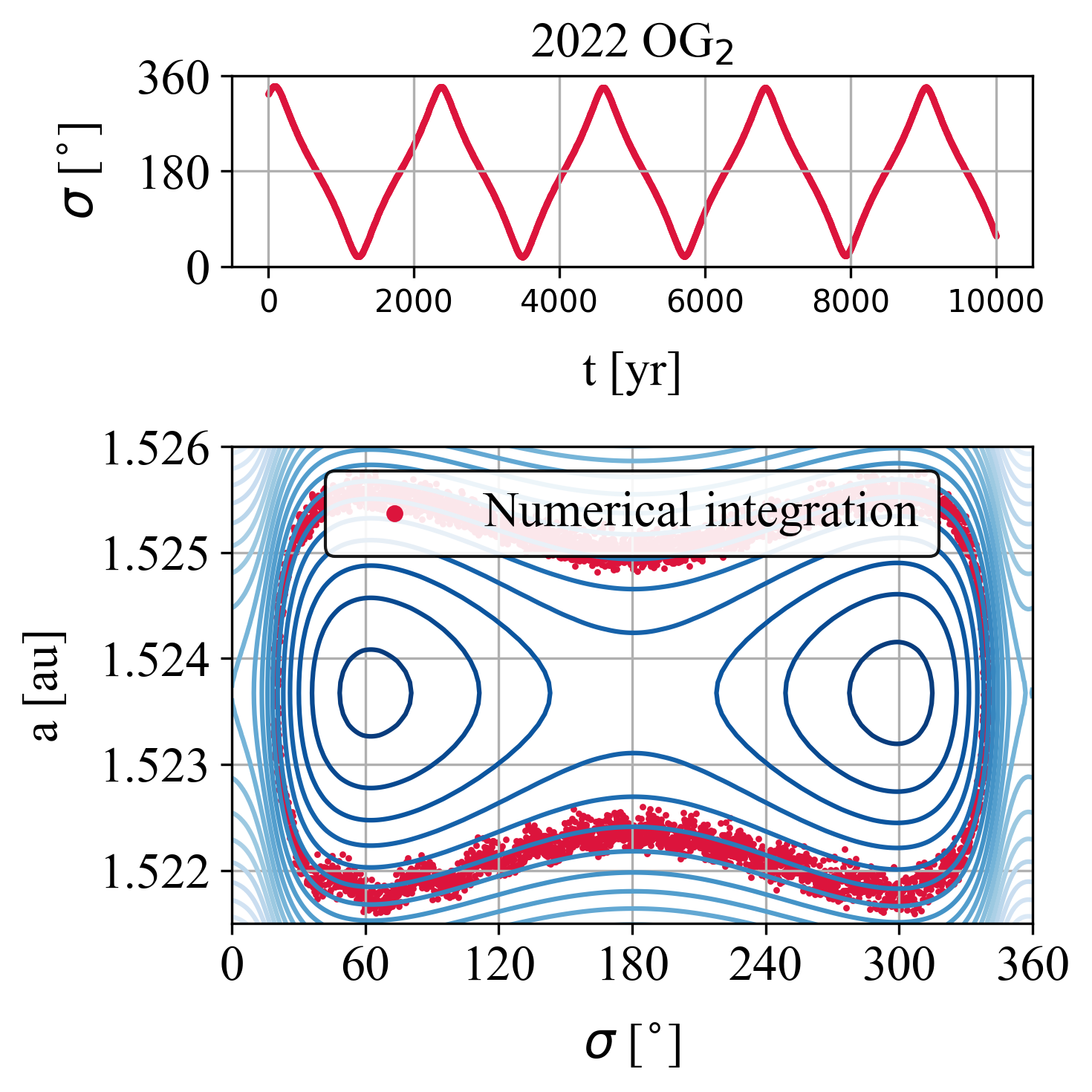}
        \caption{}
        \label{fig:objetosNuevos_hs_a}
    \end{subfigure}
    \begin{subfigure}{0.4\textwidth}
        \centering
        \includegraphics[scale=0.42]{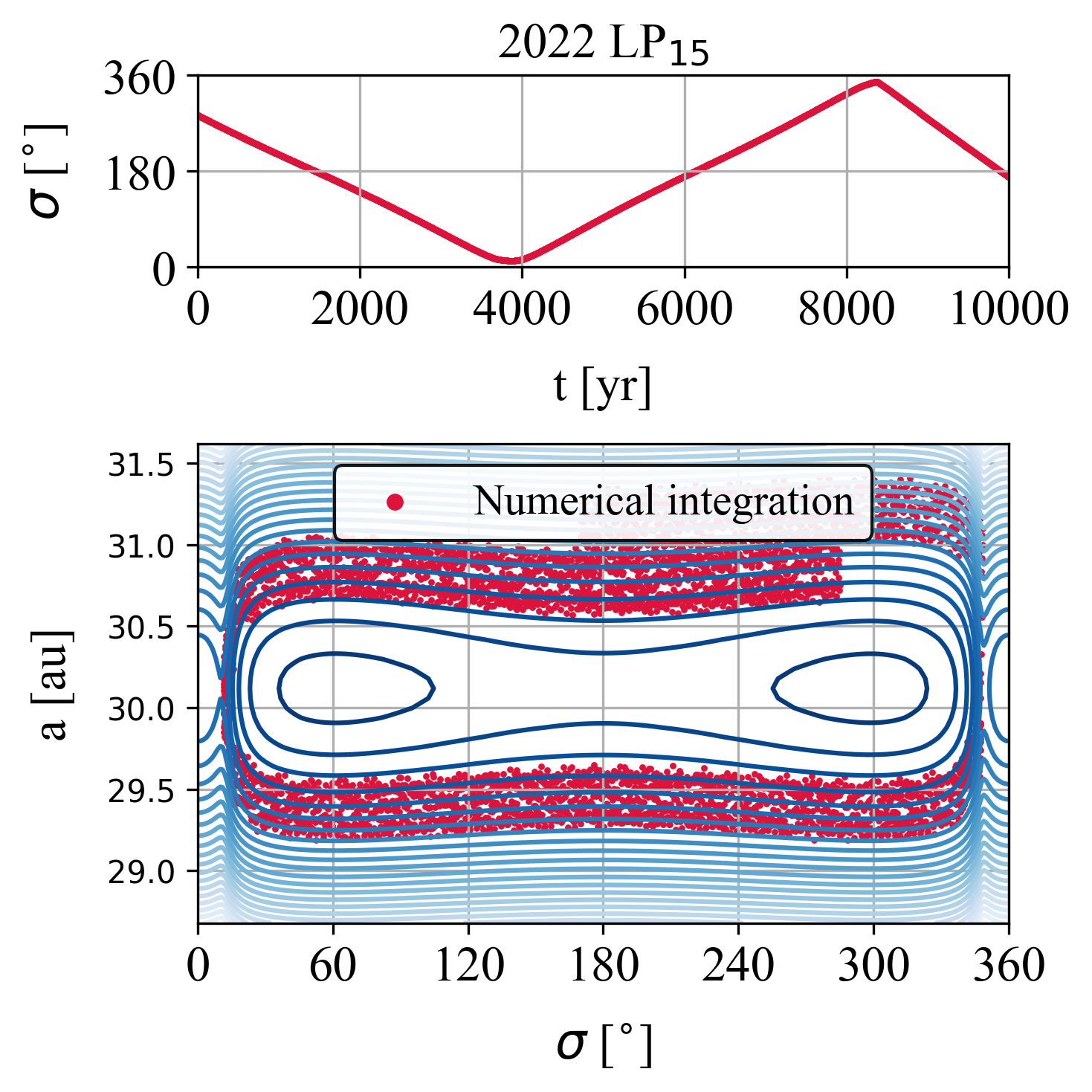}
        \caption{}
        \label{fig:objetosNuevos_hs_b}
    \end{subfigure}
    \caption{(a) Time evolution of object 2022 OG$_{2}$ ($e \sim  0.21$, $i \sim 9^{\circ}$) as an example of a Horseshoe orbit with Mars. (b) Time evolution of object 2022 LP$_{15}$ ($e \sim  0.10$, $i \sim 5^{\circ}$), an example of a Horseshoe orbit with Neptune.}
    \label{fig:objetosNuevos_hs}
\end{figure}

As we can see in Figure \ref{fig:objetosNuevos_hs}, the evolution in the phase space for Horseshoe type orbits wraps around two stable equilibrium points (L4 and L5) and one unstable equilibrium point (L3) located at $\sigma \sim 180^{\circ}$.

\subsection{Transitions and sticking}

\begin{figure}[H]
    \centering
    \begin{subfigure}{0.4\textwidth}
        \centering
        \includegraphics[scale=0.42]{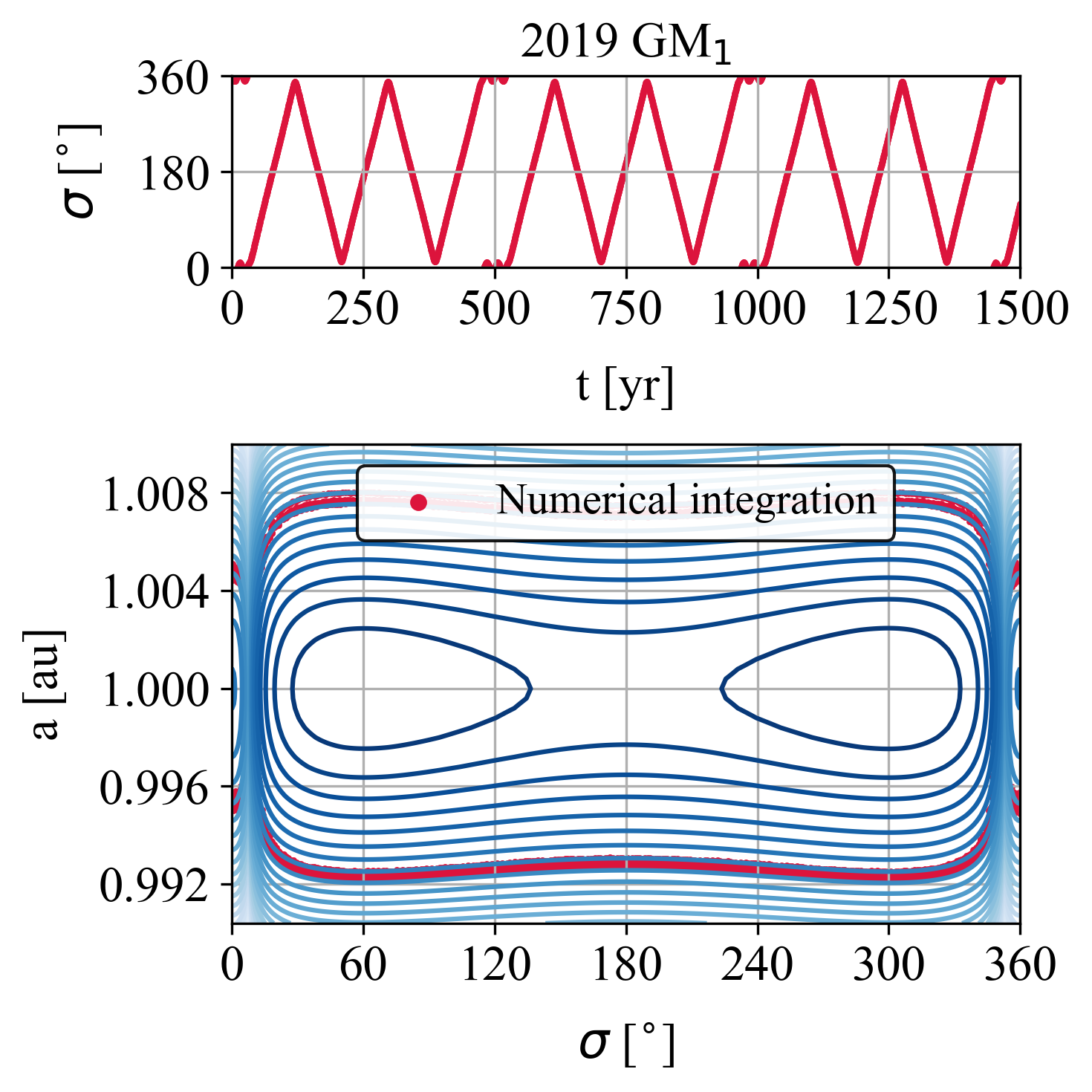}
        \caption{}
        \label{fig:otherobjectsa}
    \end{subfigure}
    \begin{subfigure}{0.4\textwidth}
        \centering
        \includegraphics[scale=0.42]{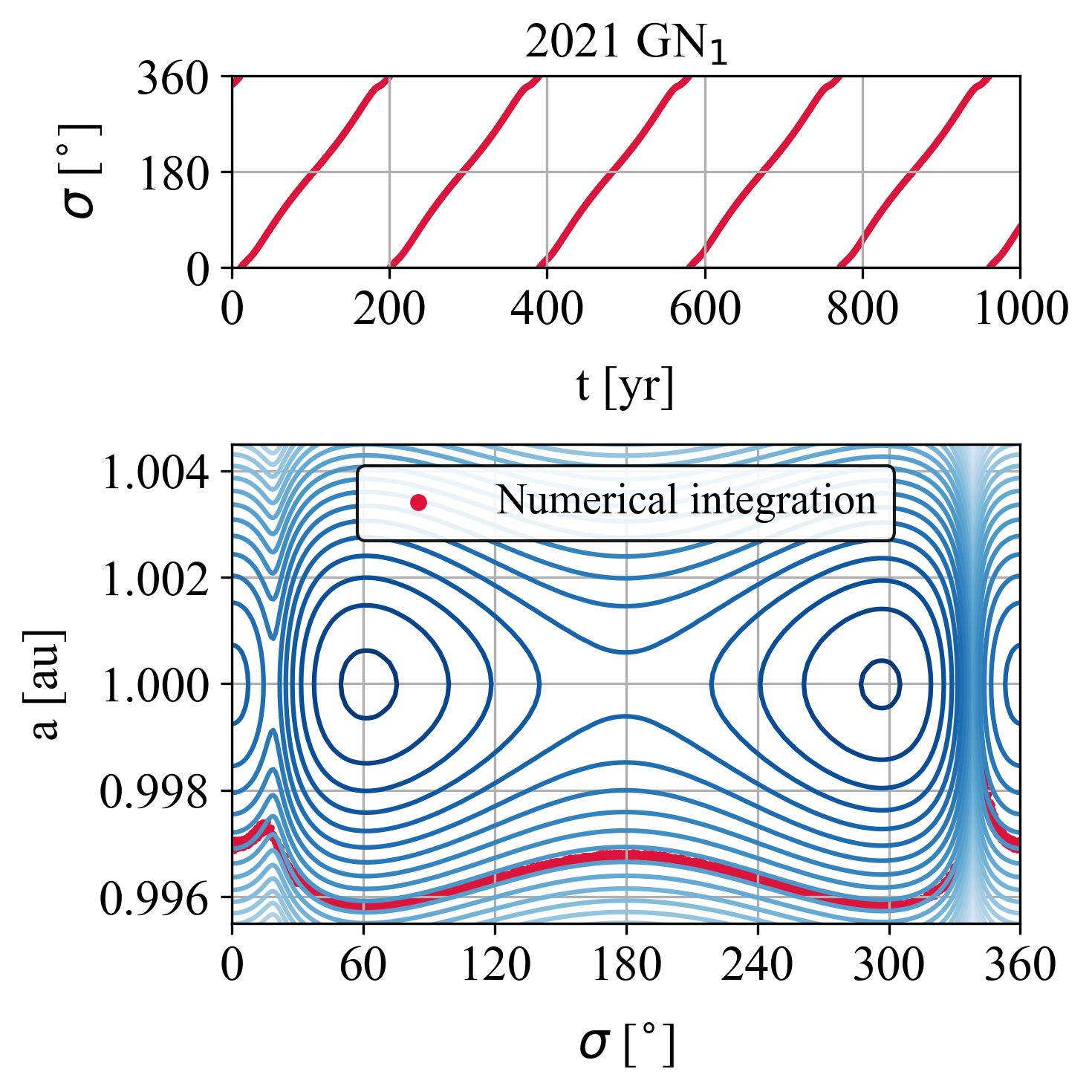}
        \caption{}
        \label{fig:otherobjectsb}
    \end{subfigure}
    \caption{(a) Time evolution of object 2019 GM$_{1}$ ($e \sim  0.07$, $i \sim 7^{\circ}$) which is in a HS-QS orbit with the Earth. (b) Time evolution of object 2021 GN$_{1}$ ($e \sim  0.19$, $i \sim 27^{\circ}$) as an example of a sticking orbit with planet Earth.}
    \label{fig:otherobjects}
\end{figure}

Another interesting orbit type in co-orbital motion are HS-QS transition orbits. They have been discussed in the literature, see for example \citet{1999Icar..137..293N}. As we can see in Figure \ref{fig:otherobjectsa} the Earth co-orbital 2019 GM$_{1}$ changes its behavior continuously between a Horseshoe and a Quasi-satellite orbit. Looking carefully, it is possible to note several short time intervals captured as Quasi-satellite with a libration period very close to the one predicted by the theory $\sim$ 22 yr.
The traditional approach to identify resonant objects relies on checking libration of the critical angle $\sigma$.
However, this criteria excludes objects whose dynamics may be significantly influenced by the resonance even if the critical angle does not exhibit libration around an equilibrium point.
We demonstrate that the orbital evolution of such objects can still be shaped by the resonance, even in the absence of strict 1:1 resonance.
In this paper we identified 25 objects showing this behavior.
Figure \ref{fig:otherobjectsb} illustrates an example of this behavior for object 2021 GN$_{1}$.

During this sticking evolution, the object's semi-major axis does not exhibit oscillations around the nominal resonant value. However, the resonant Hamiltonian clearly shows that the semi-major axis evolution is connected to the critical angle. Therefore, the dynamics is driven by the resonance.
\newline

\subsection{Current co-orbital population}
\label{subsec:coorbpop}

\begin{figure}[H]
    \centering
    \begin{subfigure}{0.4\textwidth}
        \centering
        \includegraphics[scale=0.49]{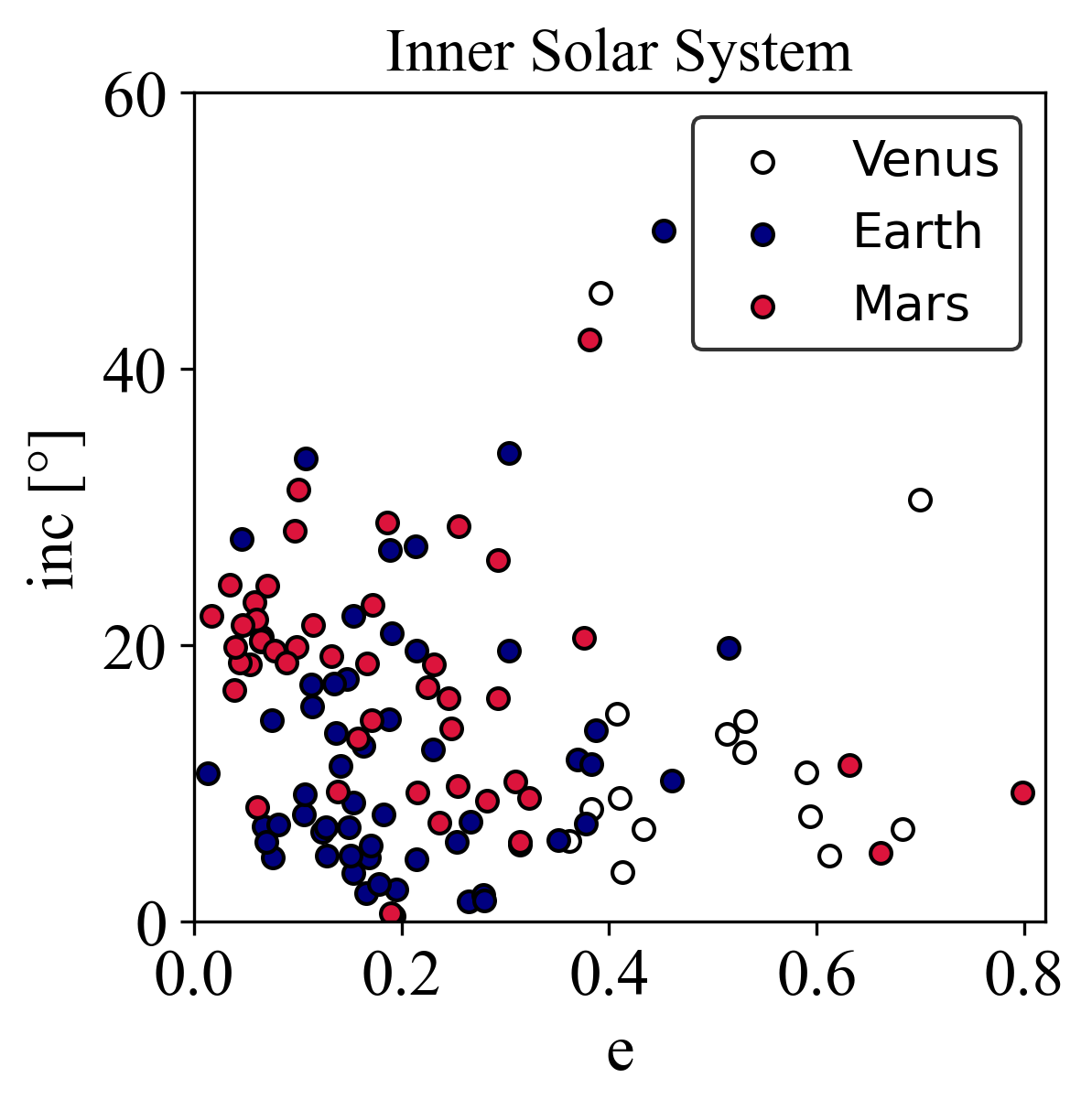}
        \caption{}
        \label{fig:catalog_a}
    \end{subfigure}
    \begin{subfigure}{0.4\textwidth}
        \centering
        \includegraphics[scale=0.49]{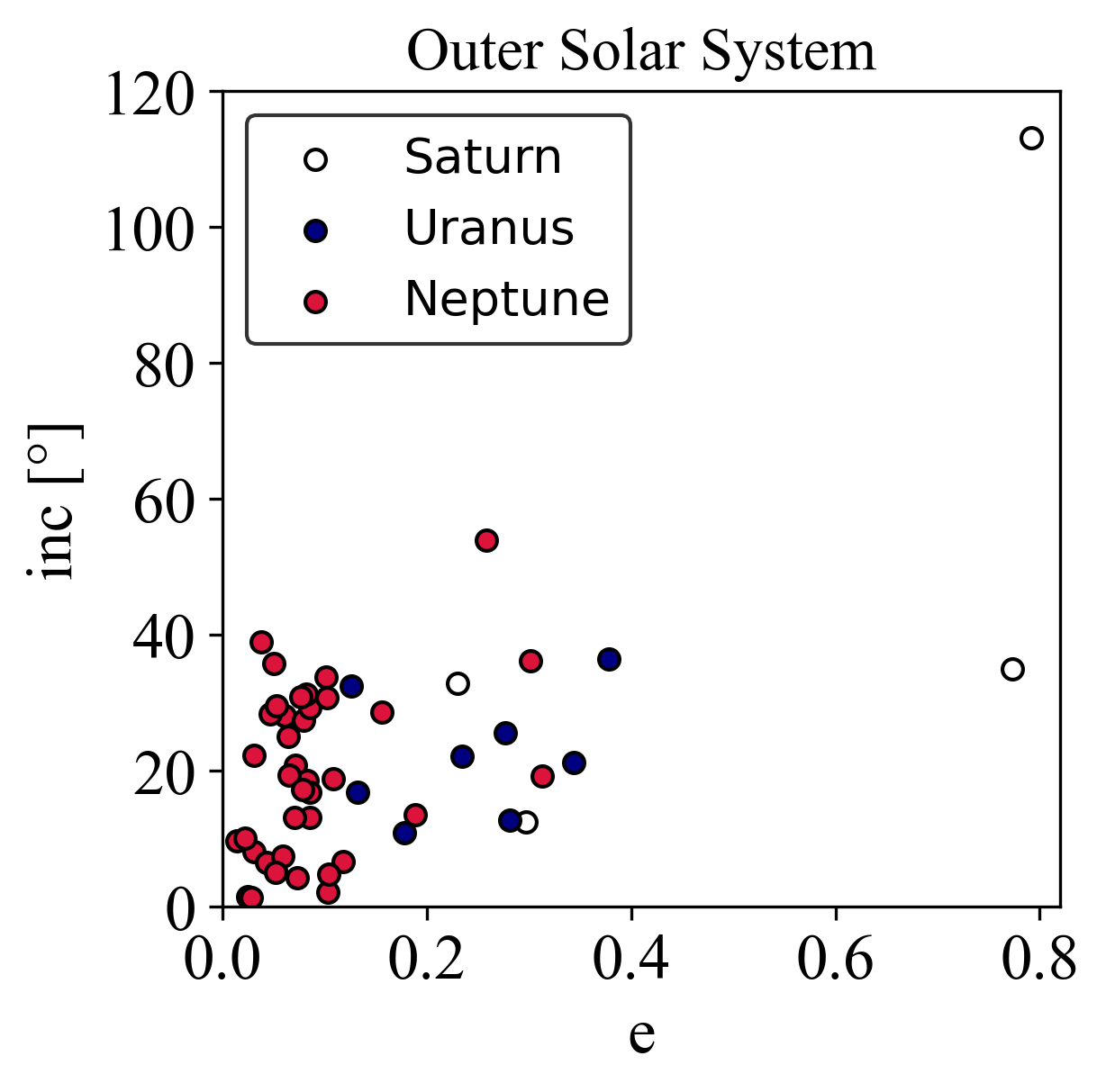}
        \caption{}
        \label{fig:catalog_b}
    \end{subfigure}
    \caption{Current catalog of Solar System co-orbitals in the $(e,i)$ plane. In different colors the different planets. See Tables \ref{tab:test_table}, \ref{tab:test_table_2} $\&$ \ref{tab:test_table_3}.}
    \label{fig:catalog}
\end{figure}

Several surveys have successfully identified new candidates for co-orbital resonant configurations within the Solar System
This has opened several questions that require further investigation to better understand the small body populations of the Solar System.
The long-term stability of these objects has been studied finding that primordial objects may only exist for Jupiter and Neptune Trojans \citep{2002Icar..160..271N}. 
Another open issue is the asymmetry between the population of Trojans. \cite{pitjeva2019masses} found that the Trojan mass ratio is $M_{L4} / M_{L5} = 1.58$ which demonstrates that more mass is trapped in L4.
Interestingly, this asymmetry is shared in the Neptune Trojan population, but the contrary is observed for Mars Trojans, despite their relatively significant population size.
A widely accepted mechanism for this asymmetry has not been proposed so far. \cite{2020A&A...642A.224M} highlighted how current formation models are not able to replicate the observed asymmetries. This discrepancy suggests that future dust trapping formation models of our Solar System should explain this. 
Figure \ref{fig:catalog} shows the distribution of all resonant co-orbital objects presented in Tables \ref{tab:test_table}, \ref{tab:test_table_2} $\&$ \ref{tab:test_table_3}.

From this plot we can see that surveys are highly biased. For instance, Venus' co-orbitals have a large eccentricity as this has been a requirement for them to be observed.
We note in Figure \ref{fig:catalog_a} that there are very few co-orbitals with quasi-circular orbits.
Also, there is a clear concentration of low-eccentricity Mars co-orbitals around $i \sim 20^{\circ}$, which has been studied in several works and proposed to be a collisional family \citep{2013Icar..224..144C, 2015Icar..252..339C, 2017MNRAS.466..489B, 2017Icar..293..243C}.
Future observational campaigns should be able to refine the orbits of several of these asteroids and confirm their dynamical classification.
We confirmed that Neptune has the second-largest group of confirmed co-orbitals in Tadpole orbits, with 34 identified.
Mars and Earth also have a large group of co-orbitals with 54 and 48 co-orbitals, respectively, while other planets have even less confirmed objects.
Mercury remains as the only planet without known co-orbitals. Additionally, certain orbits remain unoccupied by known objects, such as L5 Earth Trojans and L5 Uranus Trojans.

\newpage
\section{Conclusions}

In this work, we employ a semi-analytic resonant theory to illustrate how the 1:1 resonance equilibrium points change as a function of the asteroid's orbital elements.
The model exhibits good agreement with classical Trojan configurations.
Furthermore, studying the properties of the Hamiltonian in the full range of values of 
$(e,i)$, we have proven that the three types of co-orbital orbits: Tadpole, Quasi-satellite and Horseshoe do not exist under several resonant configurations as the equilibrium points move, merge and disappear. Additionally, we have compiled a comprehensive catalog of Solar System co-orbitals with all planets, excluding Jupiter. Overall, a total of 167 objects have been classified based on the time evolution of numerical integrations over the Hamiltonian level curves given by the theory.
Notably, this approach allowed us to discover 25 objects in non traditional resonant orbits that we call "sticking" which are characterized by a non-librating resonant angle but a clear influence of the resonance on its orbital evolution.
Our work demonstrates that phase space trajectories in ($\sigma$, $a$) plane offer a superior method to understand the resonant state of individual co-orbitals. This is applicable to other resonances as well.


\section{Declarations}

\textbf{Conflict of interest} The authors declare no conflict of interest.


\section{Acknowledgments}

We would like to thank the two anonymous reviewers for a detailed analysis of our work which allowed us to highly improve it.
We also thank to the ``Programa de Desarrollo de las Ciencias Básicas" (PEDECIBA), to ``Comisión Académica de Posgrado" (CAP), to ``Agencia Nacional de Investigación e Innovación" (ANII) and ``Comisión Sectorial de Investigación Científica" (CSIC).

\begin{appendices}

\section{GitHub Repository}
\label{app:github}

All the semi-analytical model codes were originally written in FORTRAN by T. Gallardo. 
To improve usability, we developed a Python interface that runs these base codes and displays key information for any resonance. We encourage others to use this model; the codes are publicly available in a GitHub repository:
\newline
\href{https://github.com/NicolasPan/semianalyticResonance}{github.com/NicolasPan/semianalyticResonance}
\newline
Additionally, we included some animations generated using the model and the integrations used for the orbit determination in this work.

\end{appendices}

\bibliography{biblio}

\end{document}